\documentclass[aip,reprint,amsmath,amssymb,jap,showpacs,showkeys]{revtex4-1}

\usepackage{algorithm}
\usepackage{algpseudocode}
\usepackage{graphicx}
\renewcommand{\vec}{\mathbf}
\renewcommand{\r}{\vec{r}}
\newcommand{\R}{\vec{R}}
\newcommand{\ve}{\varepsilon}
\renewcommand{\phi}{\varphi}
\renewcommand{\d}{\partial}
\newcommand{\eq}[1]{(\ref{#1})}
\newcommand{\p}{\mathrm}
\newcommand{\LL}{\mathcal{L}}

\begin{document}

\title{Elastic strain field due to an inclusion of a polyhedral shape with a~non-uniform lattice misfit}

\author{A.~V. Nenashev}
\affiliation{Rzhanov Institute of Semiconductor Physics SB RAS, 630090 Novosibirsk, Russia}
\affiliation{Novosibirsk State University, 630090 Novosibirsk, Russia}

\author{A.~V.~Dvurechenskii}
\affiliation{Rzhanov Institute of Semiconductor Physics SB RAS, 630090 Novosibirsk, Russia}
\affiliation{Novosibirsk State University, 630090 Novosibirsk, Russia}

\date{\today}

\begin{abstract}
An analytical solution in a closed form is obtained for the three-dimensional 
elastic strain distribution in an unlimited medium containing an inclusion with 
a coordinate-dependent lattice mismatch (an eigenstrain). 
Quantum dots consisting of a solid solution 
with a spatially varying composition are examples of such inclusions. 
It is assumed that both the inclusion and the surrounding medium (the matrix) 
are elastically isotropic and have the same Young modulus and Poisson ratio. 
The inclusion shape is supposed to be an arbitrary polyhedron, and the coordinate 
dependence of the lattice misfit, with respect to the matrix, 
is assumed to be a polynomial of any degree. 
It is shown that, both inside and outside the inclusion, the strain tensor is 
expressed as a sum of contributions of all faces, 
edges and vertices of the inclusion. 
Each of these contributions, as a function of the observation point's coordinates, is a product of some polynomial 
and a simple analytical function, which is 
the solid angle subtended by the face from the observation point (for a contribution of a face), 
or the potential of the uniformly charged edge (for a contribution of an edge), 
or the distance from the vertex to the observation point (for a contribution of a vertex).
The method of constructing the relevant polynomial functions is suggested. 
We also found out that similar expressions describe an electrostatic or gravitational 
potential, as well as its first and second derivatives, of a polyhedral body 
with a charge/mass density that depends on coordinates polynomially.
\end{abstract}


\maketitle

\section{Introduction}
\label{sec:introduction}

Self-assembled quantum dots are inclusions of one semiconducting material within 
another material (a matrix). Due to different lattice constants of the two 
semiconductors, such structures possess built-in elastic strain, when grown coherently. 
The strain plays a key role both in the island self-assembly during 
heteroepitaxy,\cite{Stangl2004} and in electronic properties of quantum dots due to 
the strain effect on the carrier dispersion law.\cite{Bir_Pikus_book, Van_de_Walle1989} 
The strain is especially important in type-II heterostructures with quantum dots. 
For example, the electron localization in Ge/Si quantum dots occurs in potential wells 
formed in Si matrix near the Ge inclusion due to a spatially inhomogeneous 
strain.\cite{Dvurechenskii2002nano, Zinovieva2013} The knowledge of the spatial 
distribution of the strain induced by quantum dots is, therefore, important for 
the analysis of their electronic structure. It is natural in this context that 
many theoretical works on electronic properties of epitaxial quantum 
dots\cite{Grundmann1995, Pryor1998, Stier1999, Dvurechenskii2002nano, Stoleru2002} 
begin with considerations of the strain distribution.

There exist a few inclusion shapes, for which the problem of finding 
the strain distribution has an analytical solution in a closed form. 
Examples are two-dimensional problems of inclusions having polygonal 
shapes\cite{Rodin1996, Faux1997, Glas2001, Jiang2004, Zou2011, Sun2012, Nenashev2013en, Chen2014, Yue2015, Lee2015} and some flat shapes with a curvilinear boundary;\cite{Faux1997, Zou2011} three-dimensional problems of ellipsoidal\cite{Eshelby1957, Eshelby1959, Eshelby1961, Kinoshita1971, Asaro1975, Mura1978, Kinoshita1986, Yu1994, Rahman2002} and polyhedral\cite{Chiu1977, Chiu1978, Rodin1996, Downes1997, Pearson2000, Nozaki2000, Glas2001, Li2001, Ovidko2005, Kuvshinov2008, Nenashev2010} inclusions. 
An extensive list of bibliography on this issue can be found, for example, in Refs.~\onlinecite{Ovidko2005, Kuvshinov2008, Gao2012, Wang2016}. The present work is devoted to 
a three-dimensional polygonal inclusion. This choice is motivated by the facts that 
quantum dots often have a faceted surface, and that a polyhedron is a convenient approximation to a body of an arbitrary shape.

The aim of the present study consists in finding an analytical expression 
(via elementary functions) for the three-dimensional distribution of the strain 
tensor~$\ve_{\alpha\beta}(x,y,z)$ within an inclusion (a quantum dot) and in the 
surrounding matrix, \emph{taking into account spatial inhomogeneity of the misfit strain in 
the inclusion}. We will obtain, in this paper, an analytical solution 
for an arbitrary polyhedral-shaped inclusion with the lattice misfit~$\ve_0(x,y,z)$ 
described by any polynomial function of coordinates. The lattice misfit is defined 
as a relative deviation of the unstrained lattice constant $a$ (which is determined 
by the composition of the material in a given point inside the inclusion) 
from the lattice constant of the environment $a_0$:
\begin{equation} \label{eps0}
  \ve_0(x,y,z) = \frac{a(x,y,z)-a_0}{a_0} \, .
\end{equation}
The quantity $\ve_0$ is often referred to as \emph{eigenstrain}. 
The content of epitaxial quantum dots generally represents a solid solution with 
a composition smoothly varying in space.\cite{Offermans2005, Stoffel2009, Picco2012} 
This provides the motivation for considering a coordinate-dependent lattice 
misfit~$\ve_0(x,y,z)$.

Our analysis is based on the following simplifying assumptions. The elastic strain 
is considered in the continuous-medium, isotropic approximation. The lattice 
misfit is also isotropic. The strain is small. The inclusion and the surrounding 
matrix possess the same elastic properties, i.~e., have the same values of 
Young modulus and Poisson ratio. The matrix is spread to infinity in all directions.

Let us briefly survey what is currently known about analytical expressions for the strain 
induced by inclusions with a \emph{spatially varied} lattice misfit. 
For ellipsoidal inclusions, such expressions have been known for several 
decades,\cite{Eshelby1961, Asaro1975, Rahman2002} and generalized for the case 
of anisotropic medium.\cite{Mura1978} For a two-dimensional problem, where the 
inclusion shape and composition depend on two coordinates only, analytical answers 
were also obtained in the cases of linear,\cite{Sun2012, Chen2014} 
quadratic\cite{Yue2015} and, finally, arbitrary polynomial coordinate dependence of 
the lattice misfit,\cite{Lee2015} taking into account also elastic anisotropy 
and piezoeffect.

The problem of the \emph{polyhedral} shaped inclusion with a constant lattice misfit 
was solved analytically in particular cases (a cuboid, a pyramid),\cite{Chiu1977, Chiu1978, Downes1997, Pearson2000, Glas2001, Li2001, Ovidko2005} as well as for the 
general polyhedron.\cite{Rodin1996, Nozaki2000, Kuvshinov2008, Nenashev2010} 
Kuvshinov\cite{Kuvshinov2008} showed that an analytical expression for the 
strain distribution should exist in the case of polynomial coordinate dependence 
of the lattice misfit, but he did not provide an explicit analytical formula. 
One can find, however, in the literature, the explicit solutions for a related 
problem---calculation of the Newtonian potential and its derivatives for a massive body 
of a polyhedral shape with an inhomogeneously distributed mass. 
This problem, being mathematically 
equivalent to the elasticity problem (see Section~\ref{sec:analogy}), was solved 
however only in the simplest cases of linear\cite{Pohanka1998,Holstein2003, Hamayun2009, 
DUrso2014, Conway2015} and quadratic\cite{Bhaskara-Rao1990, Gallardo-Delgado2003, 
Gokula2015} dependence of the mass density on coordinates. Thus, the task of 
the present paper has not been solved yet.

The present study is based on the approach developed in our previous work, 
Ref.~\onlinecite{Nenashev2010}. In this work, the strain due to a polyhedral inclusion 
with a \emph{constant} lattice misfit is shown to be a sum of contributions 
of \emph{faces} and \emph{edges} of the polyhedron. A contribution of a face 
is proportional to the solid angle subtended by this face. (The role of solid angles 
in calculating the potential and the strain is mentioned also 
in Refs.~\onlinecite{Werner1996, Downes1997, Kuvshinov2008}.) A contribution of 
an edge is proportional to the potential induced of this edge, as if it were 
uniformly charged (see also Ref.~\onlinecite{Werner1996}). In the present paper, 
this result is generalized to the case of a non-uniform lattice misfit in the following 
way. At first, the coefficients of proportionality at face and edge contributions 
become polynomial functions of coordinates of the observation point. At second, 
the contributions of \emph{vertices} of the polyhedron to the strain tensor 
are also introduced; for each vertex, its contribution is equal to the product 
of some polynomial function of coordinates and the distance between this vertex 
and the observation point. We prove that such modification of the results 
of Ref.~\onlinecite{Nenashev2010} indeed provides a solution for the problem 
of the present work, and suggest an algorithm for finding the coefficients 
of all needed polynomials. 

Before proceeding to the development of the solution, let us introduce some notations. 
The coordinate-dependent lattice misfit $\ve_0(\r)$ defined by Eq.~\eq{eps0} 
can be represented as
\begin{equation} \label{rho}
  \ve_0(\r) = 
  \begin{cases}
    f(\r) & \text{if } \r \text{ is inside the inclusion}, \\
	0     & \text{if } \r \text{ is outside the inclusion},
  \end{cases}
\end{equation}
with $f(\r)$ being a polynomial function of coordinates $x,y,z$ of some degree $N$:
\begin{equation} \label{f-via-C}
  f(x,y,z) = \sum\limits_{\substack{m,n,p \\ m+n+p \leq N}} C_{mnp} \, x^m y^n z^p .
\end{equation}
Sometimes it is convenient to represent the function $\ve_0(\r)$ as
\begin{equation} \label{rho-via-chi}
  \ve_0(\r) = f(\r) \; \chi(\r),
\end{equation}
where $\chi(\r)$ is the characteristic function that
determines whether the point $\r$ belongs to the inclusion:
\begin{equation} \label{chi}
  \chi(\r) = 
  \begin{cases}
    1 & \text{if } \r \text{ is inside the inclusion}, \\
	0 & \text{otherwise}.
  \end{cases}
\end{equation}

Our goal is to calculate the strain tensor $\ve_{\alpha\beta}$ 
at an arbitrary point $\R=(X,Y,Z)$:
\begin{equation} \label{what-is-epsilon}
  \ve_{\alpha\beta} (\R)  =  \; ?
\end{equation}

Also, as we will see in Section~\ref{sec:analogy}, an important function 
for our consideration is the electrostatic potential $\phi(\r)$ that 
is produced by the charge distribution defined by the function $\ve_0(\r)$. 
More formally, $\phi(\r)$ is the solution of Poisson equation
\begin{equation} \label{phi-Poisson}
  \Delta \phi(\r) = -4 \pi \ve_0(\r) ,
\end{equation}
which vanishes at infinity. We will also represent the analytical 
expressions for the potential $\phi(\r)$ and its first and second derivatives. 

This paper is organized as follows. The connection between the strain~$\ve_{\alpha\beta}$ 
and the potential~$\phi$ is explained in Section~\ref{sec:analogy}. 
In Sections~\ref{sec:monoms} and~\ref{sec:primitives} we will express the potential~$\phi$ 
in terms of more simple quantities---four kinds of ``primitives'' introduced in 
Section~\ref{sec:primitives}. They can be considered as potentials induced by a charge 
inhomogeneously distributed over the inclusion, or over some its face, or over some its edge, 
or by a double layer inhomogeneously covering one of the inclusion faces. 
Then, in Sections~\ref{sec:reducing} and~\ref{sec:L} we will show how to find analytical expressions 
for the ``primitives''. All these results are summed up in Section~\ref{sec:main}, 
yielding the final analytical formula for the strain distribution $\ve_{\alpha\beta}(\R)$ 
(Subsection~\ref{sec:main-expression}) and the algorithm for finding the polynomial coefficients 
contributing to this formula (Subsection~\ref{sec:main-algorithm}). 
Analytical results for the potential $\phi$ and its derivatives are also supplied. 
Section~\ref{sec:examples} provides some examples of applying the general answer to particular cases. 
Concluding remarks are presented in Section~\ref{sec:conclusions}.

\section{Elastic-electrostatic analogy}
\label{sec:analogy}

The displacement field $\vec{u}(\r)$ in an isotropic elastic medium caused by a point-like isotropic inclusion (put at $\r=0$) looks just like the electric field of a point charge: 
\begin{equation} 
  \vec{u}(\r) \propto \frac{\r}{r^3} .
\end{equation}
This simple fact provides a basis for the analogy between the problems of electrostatics 
and of elasticity theory. More precisely, if the elastic medium is infinite and isotropic, 
and its Young module and Poisson ratio are constant throughout the whole space, then the displacement 
induced by the eigenstrain distribution $\ve_0(\r)$ can be expressed as follows\cite{Landau_book_elasticity, Davies1998}:
\begin{equation} \label{u-via-potential}
  \vec{u} (\r)  =  - \Lambda \, \frac{\d\phi(\r)}{\d\r}  \, ,
\end{equation}
where $\phi(\r)$ is the potential defined by Eq.~\eq{phi-Poisson}, and 
the coefficient $\Lambda$ is related to the Poisson ratio $\nu$:
\begin{equation} \label{Lambda-via-nu}
  \Lambda  =  \frac{1}{4\pi} \, \frac{1+\nu}{1-\nu} \, .
\end{equation}

Since the strain tensor $\ve_{\alpha\beta}$ is defined via derivatives of the displacement,
\begin{equation} \label{eps-via-u}
  \ve_{\alpha\beta} (\r)  
  =  \frac12 \left( \frac{\d u_\alpha}{\d r_\beta} + \frac{\d u_\beta}{\d r_\alpha} \right) 
  - \delta_{\alpha\beta} \, \ve_0(\r)  \, ,
\end{equation}
then it is a combination of \emph{second} derivatives of $\phi$:\cite{Davies1998, Nenashev2010}
\begin{equation} \label{epsilon-via-potential}
  \ve_{\alpha\beta} (\r)  =  - \Lambda \, \frac{\d^2\phi(\r)}{\d r_\alpha \, \d r_\beta} 
  - \delta_{\alpha\beta} \, \ve_0(\r) .
\end{equation}
Here $\delta_{\alpha\beta}$ is the Kronecker delta.

Obviously, the potential $\phi$ has an integral representation:
\begin{equation} \label{potential-integral}
  \phi(\R)  
  =  \iiint \ve_0(\r) \, \frac{d^3 r}{|\r-\R|}  
  \equiv \iiint\limits_{\text{inclusion}} f(\r) \, \frac{d^3 r}{|\r-\R|}  \, ,
\end{equation}
as well as its first and second derivatives:
\begin{equation} \label{potential-deriv-integral}
  \frac{\d\phi(\R)}{\d\R}  =  \iiint \frac{\d\ve_0(\r)}{\d\r} \, \frac{d^3 r}{|\r-\R|} \, ,
\end{equation}
\begin{equation} \label{potential-2deriv-integral}
  \frac{\d^2\phi(\R)}{\d R_\alpha \d R_\beta}  
  =  \iiint \frac{\d^2\ve_0(\r)}{\d r_\alpha \d r_\beta} \, \frac{d^3 r}{|\r-\R|} \, .
\end{equation}
(Triple integrals without specification of the domain are assumed to be over the whole space.) 
The latter equation provides also an integral representation for the strain tensor, 
by virtue of Eq.~\eq{epsilon-via-potential}. 
It is important to note that integrals~\eq{potential-integral}--\eq{potential-2deriv-integral} 
were extensively studied in the context of geophysical\cite{Nagy1966, Paul1974, Okabe1979, Pohanka1988, Holstein1996, Pohanka1998, Bhaskara-Rao1990, Holstein2002, Holstein2003, Gallardo-Delgado2003, Garcia-Abdeslem2005, Hamayun2009, DUrso2014, Gokula2015, DUrso2015} and astronomy/spacecraft\cite{Werner1994, Werner1996, Broucke1999, Conway2015} problems. 
But, as mentioned in Introduction, fully analytical expressions for polyhedral bodies were obtained 
only when the density $f(\r)$ is a constant ($N$=0), a linear ($N$=1) or a quadratic ($N$=2) function 
of coordinates.

In the next Section, we transform these integrals into a form more convenient for further analysis.

\section{From polynomials to monomials}
\label{sec:monoms}

It is easier to deal with potentials induced by charge density distributions of the following form:
\begin{equation} \label{rho-mnp}
  \rho_{mnp}(\r;\R) = (x-X)^m (y-Y)^n (z-Z)^p \, \chi(\r) ,
\end{equation}
where $m,n,p$ are non-negative integer numbers, 
and the characteristic function $\chi(\r)$ is defined in Eq.~\eq{chi}.
One can expand the ``actual'' density $\ve_0(\r)$ into a series of terms $\rho_{mnp}$:
\begin{equation} \label{rho-via-rho-mnp}
  \ve_0(\r) = \sum\limits_{\substack{m,n,p \\ m+n+p \leq N}} \tilde C_{mnp}(\R) \; \rho_{mnp}(\r;\R) .
\end{equation}
Each of the coefficients $\tilde C_{mnp}$ is a polynomial of $X,Y,Z$ of degree $N-m-n-p$. 
The explicit expression for polynomials $\tilde C_{mnp}(\R)$ can be obtained by substituting 
Eqs.~\eq{rho}, \eq{f-via-C} and~\eq{rho-mnp} into Eq.~\eq{rho-via-rho-mnp}, that leads to 
the identity
\begin{multline} \label{C-via-tilde-C}
  \sum\limits_{m,n,p} C_{mnp} \, x^m y^n z^p \\
  = \sum\limits_{m,n,p} \tilde C_{mnp}(\R) \; (x-X)^m (y-Y)^n (z-Z)^p ,
\end{multline}
whence
\begin{multline} \label{tilde-C}
  \tilde C_{mnp}(\R) = \\
  \sum\limits_{\substack{m' \geqslant m \\ n' \geqslant n \\ p' \geqslant p}} C_{m'n'p'} 
  \binom{m'}{m} \binom{n'}{n} \binom{p'}{p} X^{m'-m} Y^{n'-n} Z^{p'-p} ,
\end{multline}
where symbols $\binom{m'}{m}$ are binomial coefficients.

Substituting the expansion~\eq{rho-via-rho-mnp} 
into Eqs.~\eq{potential-integral}--\eq{potential-2deriv-integral}, 
one can get similar expansions for the potential $\phi(\R)$ and its derivatives:
\begin{equation} \label{phi-via-phi-mnp}
  \phi(\R) = \sum\limits_{m,n,p} \tilde C_{mnp}(\R) \; \phi_{mnp}(\R) ,
\end{equation}
\begin{equation} \label{phi-deriv-via-phi-mnp}
  \frac{\d\phi(\R)}{\d R_\alpha} = \sum\limits_{m,n,p} \tilde C_{mnp}(\R) \; \phi^{(mnp)}_{,\alpha}(\R) ,
\end{equation}
\begin{equation} \label{phi-2deriv-via-phi-mnp}
  \frac{\d^2\phi(\R)}{\d R_\alpha \d R_\beta} 
  = \sum\limits_{m,n,p} \tilde C_{mnp}(\R) \; \phi^{(mnp)}_{,\alpha\beta}(\R) ,
\end{equation}
where symbols $\phi_{mnp}$, $\phi^{(mnp)}_{,\alpha}$, $\phi^{(mnp)}_{,\alpha\beta}$ denote potentials and 
their derivatives induced by ``monomial'' charge distributions $\rho_{mnp}$ 
defined in Eq.~\eq{rho-mnp}:
\begin{equation} \label{phi-mnp}
  \phi_{mnp}(\R) = \iiint (x-X)^m (y-Y)^n (z-Z)^p \, \chi(\r) \, \frac{d^3 r}{|\r-\R|} \, ,
\end{equation}
\begin{multline} \label{phi-mnp-alpha}
  \phi^{(mnp)}_{,\alpha}(\R) \\
  = \iiint \frac{\d \big[ (x-X)^m (y-Y)^n (z-Z)^p \, \chi(\r) \big]}{\d r_\alpha} \; 
  \frac{d^3 r}{|\r-\R|} \, ,
\end{multline}
\begin{multline} \label{phi-mnp-alpha-beta}
  \phi^{(mnp)}_{,\alpha\beta}(\R) \\
  = \iiint \frac{\d^2 \big[ (x-X)^m (y-Y)^n (z-Z)^p \, \chi(\r) \big]}{\d r_\alpha \d r_\beta} \; 
  \frac{d^3 r}{|\r-\R|} \, .
\end{multline}

\section{Four primitives $\varphi_{mnp}$, $\Phi^{(i)}_{mnp}$, $\Omega^{(i)}_{mnp}$, $L^{(k)}_{mnp}$}
\label{sec:primitives}

\begin{figure}
  \includegraphics[width=\columnwidth]{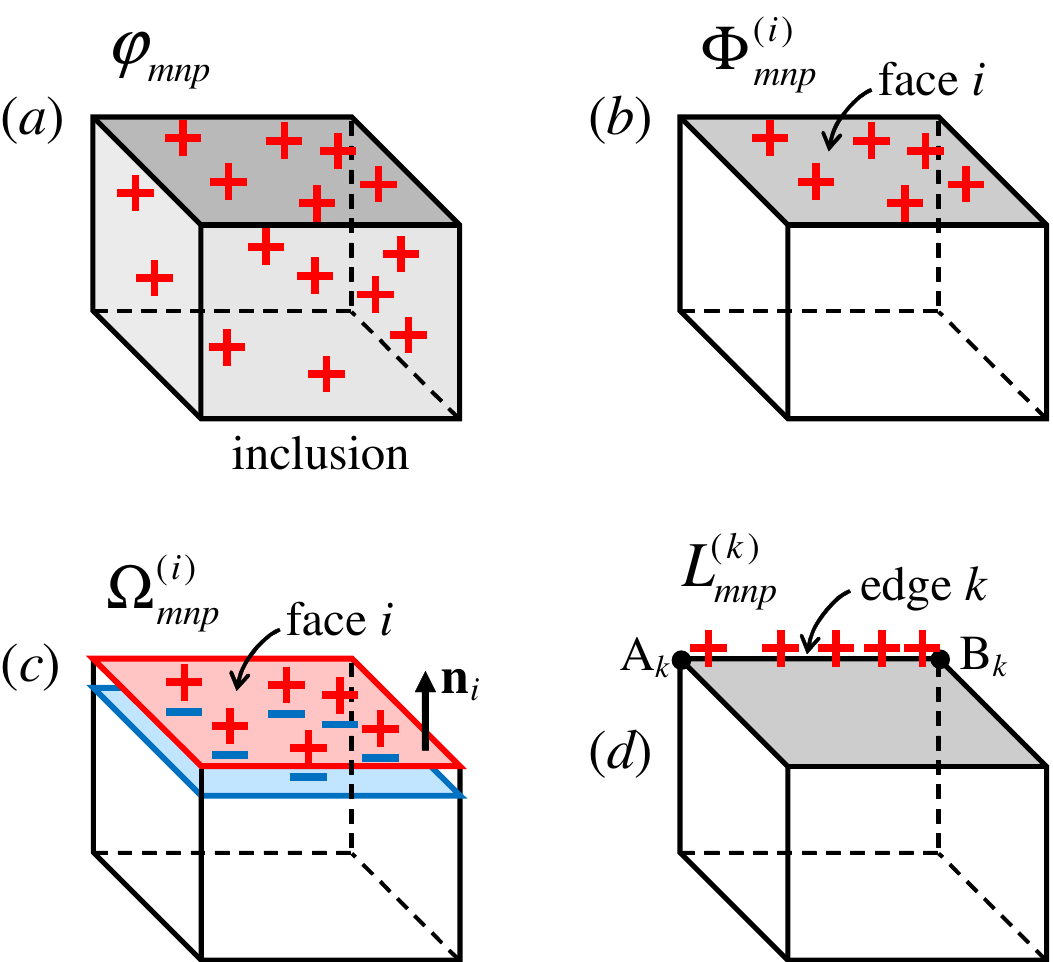}
  \caption{The primitives $\varphi_{mnp}$, $\Phi^{(i)}_{mnp}$, $\Omega^{(i)}_{mnp}$, $L^{(k)}_{mnp}$ 
  as potentials induced at point $\R=(X,Y,Z)$ by the following inhomogeneous charge distributions 
  as functions of coordinates $x,y,z$: 
  \emph{(a)} by a charged inclusion with \emph{volume} density $(x-X)^m (y-Y)^n (z-Z)^p$; 
  \emph{(b)} by charged $i$th face with \emph{surface} charge density $(x-X)^m (y-Y)^n (z-Z)^p$; 
  \emph{(c)} by $i$th face covered by a doubly charged layer with \emph{surface density of dipole moment} $(x-X)^m (y-Y)^n (z-Z)^p$; 
  \emph{(d)} by $k$th edge with \emph{linear} charge density $(x-X)^m (y-Y)^n (z-Z)^p$.}
  \label{fig:primitives}
\end{figure}

We will consider four types of ``primitive constituents'' of the potential and its derivatives. 
The first one, $\varphi_{mnp}$, is the potential 
provided at point $\R=(X,Y,Z)$ by the charged inclusion (Fig.~\ref{fig:primitives}a) with the charge density depending on 
coordinates $x,y,z$ as $(x-X)^m (y-Y)^n (z-Z)^p$:
\begin{equation} \label{phi-mnp-def}
  \varphi_{mnp} (\R) = \iiint\limits_{\text{inclusion}} \frac{(x-X)^m (y-Y)^n (z-Z)^p}{|\r-\R|} \, dV .
\end{equation}
This expression is just a different form of Eq.~\eq{phi-mnp}.

The second primitive, $\Phi^{(i)}_{mnp}$, is the potential of one of the inclusion faces (number $i$) 
charged with surface charge density $(x-X)^m (y-Y)^n (z-Z)^p$ (Fig.~\ref{fig:primitives}b):
\begin{equation} \label{Phi-i-mnp-def}
  \Phi^{(i)}_{mnp} (\R) = \iint\limits_{\text{face} \, i} \frac{(x-X)^m (y-Y)^n (z-Z)^p}{|\r-\R|} \, dS .
\end{equation}

The third primitive, $\Omega^{(i)}_{mnp}$, is the potential of one of $i$th face 
covered with a thin dipole layer with dipole moment density $(x-X)^m (y-Y)^n (z-Z)^p$:
\begin{equation} \label{Omega-i-mnp-def}
  \Omega^{(i)}_{mnp} (\R) 
  = \iint\limits_{\text{face} \, i} (x-X)^m (y-Y)^n (z-Z)^p \, 
  \frac{(\R-\r) \cdot \vec{n}_i}{|\r-\R|^3} \, dS ,
\end{equation}
where $\vec{n}_i$ is the unit normal vector to $i$th face, directed outside the inclusion 
(see Fig.~\ref{fig:primitives}c). 

And the fourth one, $L^{(k)}_{mnp}$, is the potential due to $k$th edge 
of the inclusion, being charged with linear charge density $(x-X)^m (y-Y)^n (z-Z)^p$:
\begin{equation} \label{L-k-mnp-def}
  L^{(k)}_{mnp} (\R) 
  = \int\limits_{\text{A}_k}^{\text{B}_k}  \frac{(x-X)^m (y-Y)^n (z-Z)^p}{|\r-\R|} \, dl ,
\end{equation}
where $\text{A}_k$ and $\text{B}_k$ are two ends of the $k$th edge (Fig.~\ref{fig:primitives}d).

Significance of these primitives for our study consists in the fact that the quantities 
$\phi_{mnp}$, $\phi^{(mnp)}_{,\alpha}$, $\phi^{(mnp)}_{,\alpha\beta}$ that appear 
in Eqs.~\eq{phi-via-phi-mnp}--\eq{phi-2deriv-via-phi-mnp} can be expressed 
through them. As shown in Appendix~\ref{app:chi}, $\phi^{(mnp)}_{,\alpha}$ has the representation
\begin{equation} \label{phi-alpha-via-phi-Phi}
  \phi^{(mnp)}_{,\alpha}  =  
  \phi_{(mnp),\alpha} 
  - \sum\limits_{\substack{i \\ \text{(faces)}}} n_{i\alpha} \Phi^{(i)}_{mnp} \, ,
\end{equation}
where $\phi_{(mnp),\alpha}$ is the following volume integral:
\begin{multline} \label{phi-low-mnp-alpha}
  \phi_{(mnp),\alpha} \\
  = \iiint \frac{\d \big[ (x-X)^m (y-Y)^n (z-Z)^p \big]}{\d r_\alpha} \, \chi(\r) \; 
  \frac{d^3 r}{|\r-\R|} \, .
\end{multline}
One can conclude from comparison between Eqs.~\eq{phi-mnp} and~\eq{phi-low-mnp-alpha} that 
\begin{subequations} \label{phi-mnp-low-calc}
\begin{eqnarray} 
  \phi_{(mnp),x} &=& m \, \phi_{m-1,n,p} \, , \\
  \phi_{(mnp),y} &=& n \, \phi_{m,n-1,p} \, ,  \\
  \phi_{(mnp),z} &=& p \, \phi_{m,n,p-1} \, .  
\end{eqnarray}
\end{subequations}

\begin{figure}
  \includegraphics[width=\columnwidth]{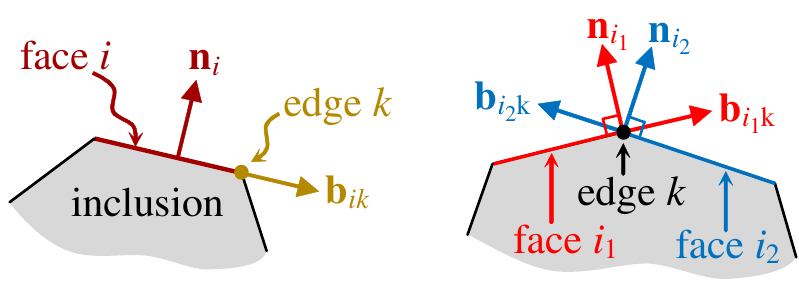}
  \caption{Illustration of the definition of unit vectors $\vec{n}_{i}$ and $\vec{b}_{ik}$ (left); 
  four unit vectors contributing to the definition~\eq{lambda-k-alpha-beta} of the tensor 
  $\lambda^{(k)}_{\alpha\beta}$ (right).}
  \label{fig:n-b}

\end{figure}

In order to write down a representation of $\phi^{(mnp)}_{,\alpha\beta}$ in terms of the primitives, 
let us introduce, for each face $i$ and each edge $k$ adjacent to this face, a unit vector $\vec{b}_{ik}$ 
directed perpendicularly to $k$th edge, along $i$th face and out of this face (see Fig.~\ref{fig:n-b}). 
Also, for each edge $k$, we define a tensor $\lambda^{(k)}_{\alpha\beta}$ as follows:\cite{Nenashev2010}
\begin{equation} \label{lambda-k-alpha-beta}
  \lambda^{(k)}_{\alpha\beta} 
  = n_{i_1\alpha} b_{i_1k\beta} 
  + n_{i_2\alpha} b_{i_2k\beta} \, ,
\end{equation}
where $i_1$ and $i_2$ are numbers of the two faces, which intersect at $k$th edge, 
as illustrated in Fig.~\ref{fig:n-b}. 
Then, the quantity $\phi^{(mnp)}_{,\alpha\beta}$ takes the following representation 
(for derivation, see Appendix~\ref{app:chi}):
\begin{multline} \label{phi-alpha-beta-via-primitives}
  \phi^{(mnp)}_{,\alpha\beta}  
  =  \phi_{(mnp),\alpha\beta} \\
  + \sum\limits_{\substack{i \\ \text{(faces)}}} \Big( 
    - n_{i\alpha} \Phi^{(i)}_{(mnp),\beta} 
	- n_{i\beta} \Phi^{(i)}_{(mnp),\alpha} 
	+ n_{i\alpha} n_{i\beta} n_{i\gamma} \Phi^{(i)}_{(mnp),\gamma}
  \Big) \\
  + \sum\limits_{\substack{i \\ \text{(faces)}}} n_{i\alpha} n_{i\beta} \Omega^{(i)}_{mnp} 
  + \sum\limits_{\substack{k \\ \text{(edges)}}} \lambda^{(k)}_{\alpha\beta} L^{(k)}_{mnp} \, ,
\end{multline}
where $\phi_{(mnp),\alpha\beta}$ and $\Phi^{(i)}_{(mnp),\alpha}$ are defined as follows:
\begin{multline} \label{phi-low-mnp-alpha-beta}
  \phi_{(mnp),\alpha\beta} \\
  = \iiint \frac{\d^2 \big[ (x-X)^m (y-Y)^n (z-Z)^p \big]}{\d r_\alpha \d r_\beta} \, \chi(\r) \; 
  \frac{d^3 r}{|\r-\R|} \, ,
\end{multline}
\begin{multline} \label{Phi-mnp-alpha}
  \Phi^{(i)}_{(mnp),\alpha} \\
  = \iint\limits_{\text{face} \, i} 
  \frac{\d \big[ (x-X)^m (y-Y)^n (z-Z)^p \big]}{\d r_\alpha} \; \frac{dS}{|\r-\R|} \, .
\end{multline}
From the comparison between Eqs.~\eq{phi-mnp} and~\eq{phi-low-mnp-alpha-beta} 
it is easy to express the quantities $\phi_{(mnp),\alpha\beta}$ through primitives $\phi_{m'n'p'}$:
\begin{subequations} \label{phi-mnp-low2-calc}
\begin{eqnarray} 
  \phi_{(mnp),xx} &=& m(m-1) \, \phi_{m-2,n  ,p} \, , \\
  \phi_{(mnp),xy} &=& mn     \, \phi_{m-1,n-1,p} \, , 
\end{eqnarray}
\end{subequations}
and so on. Similarly, the comparison between Eqs.~\eq{Phi-i-mnp-def} and~\eq{Phi-mnp-alpha} provides the following relations:
\begin{subequations} \label{Phi-mnp-low-calc}
\begin{eqnarray} 
  \Phi^{(i)}_{(mnp),x} &=& m \, \Phi^{(i)}_{m-1,n,p} \, , \\
  \Phi^{(i)}_{(mnp),y} &=& n \, \Phi^{(i)}_{m,n-1,p} \, ,  \\
  \Phi^{(i)}_{(mnp),z} &=& p \, \Phi^{(i)}_{m,n,p-1} \, .  
\end{eqnarray}
\end{subequations}

In the simplest case of $m=n=p=0$, Eq.~\eq{phi-alpha-beta-via-primitives} 
is simplified to\cite{Werner1996,Nenashev2010}
\begin{equation} \label{phi-000-alpha-beta-via-primitives}
  \phi^{(000)}_{,\alpha\beta}  
  =  \sum\limits_i n_{i\alpha} n_{i\beta} \Omega^{(i)}_{000} 
  + \sum\limits_k \lambda^{(k)}_{\alpha\beta} L^{(k)}_{000} \, .
\end{equation}
The primitives $\Omega^{(i)}_{000}$ and $L^{(k)}_{000}$ 
have simple physical and geometrical meanings and 
well-known analytical representations. Below we will drop the index ``000'' at them:
\begin{equation} 
  \Omega^{(i)}_{000} \equiv \Omega^{(i)} , \quad
  L^{(k)}_{000} \equiv L^{(k)} .
\end{equation}

\begin{figure}
  \includegraphics[width=0.8\columnwidth]{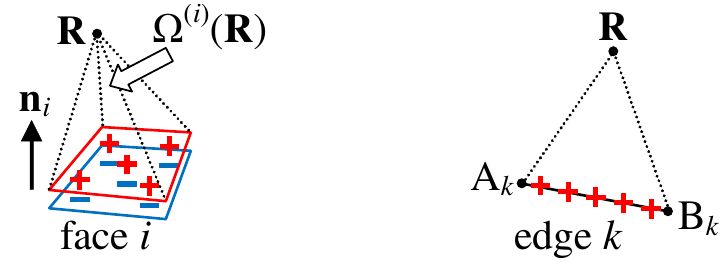}
  \caption{The meaning of the quantity $\Omega^{(i)}(\R)$ as a potential 
  due to a dipole layer at point $\R$ and as a solid angle (left);
  the quantity $L^{(k)}(\R)$ as a potential of a uniformly charged rod at point $\R$ (right).}
  \label{fig:Omega-L}
\end{figure}

The quantity $\Omega^{(i)}(\R)$ is the potential at point $\R$ due to a flat polygon 
($i$th face of the inclusion) 
covered by a uniform dipole layer of unit density, see Fig.~\ref{fig:Omega-L}. 
The absolute value of $\Omega^{(i)}(\R)$ 
is equal to the solid angle, at which the polygon is seen from 
the point $\R$.\cite{Tamm_book,Stratton_book} 
And the sign of $\Omega^{(i)}(\R)$ is positive if the normal vector $\mathbf{n}_i$ 
is directed towards the point $\R$, and negative otherwise. 
There are several recipes in the literature, how to calculate such solid angles 
analytically.\cite{Oosterom1983,Werner1996,Nenashev2007arxiv} 

It is important to note that the characteristic function of the inclusion $\chi(\r)$ 
can be expressed in terms of solid angles:
\begin{equation} \label{sum-Omega}
  \chi(\r) = -\frac{1}{4\pi} \sum\limits_i \Omega^{(i)}(\r) .
\end{equation}
Indeed, if the point $\r$ is inside the inclusion, then each of the quantities 
$\Omega^{(i)}(\r)$ is negative, so their sum is equal to $-4\pi$ (the total solid angle). 
When $\r$ is outside the inclusion, then some of quantities 
$\Omega^{(i)}(\r)$ are positive, and some are negative; moreover, the sum of negative 
$\Omega$'s exactly compensates the sum of positive ones. 
From Eqs.~\eq{rho-via-chi} and~\eq{sum-Omega} one can get a representation of the 
lattice misfit $\ve_0$:
\begin{equation} \label{rho-via-Omega}
  \ve_0(\r) = - \frac{f(\r)}{4\pi} \sum\limits_i \Omega^{(i)}(\r) .
\end{equation}

The quantity $L^{(k)}(\R)$ is the potential at point $\R$ of a thin, 
uniformly charged rod ($k$th edge) 
with unit linear charge density, see Fig.~\ref{fig:Omega-L}. 
Elementary integration provides the following expression 
for $L^{(k)}$:\cite{Werner1996,Nenashev2010}
\begin{equation} \label{L-000-answer}
  L^{(k)}(\R) = \log \frac
  {\p{RA}_k + \p{RB}_k + \p{A}_k\p{B}_k}
  {\p{RA}_k + \p{RB}_k - \p{A}_k\p{B}_k} \, ,
\end{equation}
where $\p{RA}_k$, $\p{RB}_k$ and $\p{A}_k\p{B}_k$ are distances between the point~$\R$ 
and two end points $\p{A}_k$ and $\p{B}_k$ of $k$th edge.

The main results of this Section are equations~\eq{phi-alpha-via-phi-Phi}, 
\eq{phi-mnp-low-calc}, \eq{phi-alpha-beta-via-primitives}, 
\eq{phi-mnp-low2-calc}, \eq{Phi-mnp-low-calc}, which (together with the results of 
the previous sections) allow one to express the strain tensor, the potential and derivatives 
of the potential via four kinds of primitives.

\section{Reducing primitives 
$\varphi_{mnp}$, $\Phi^{(i)}_{mnp}$, $\Omega^{(i)}_{mnp}$ 
to solid angles and line integrals}
\label{sec:reducing}

\begin{figure}
  \includegraphics[width=\columnwidth]{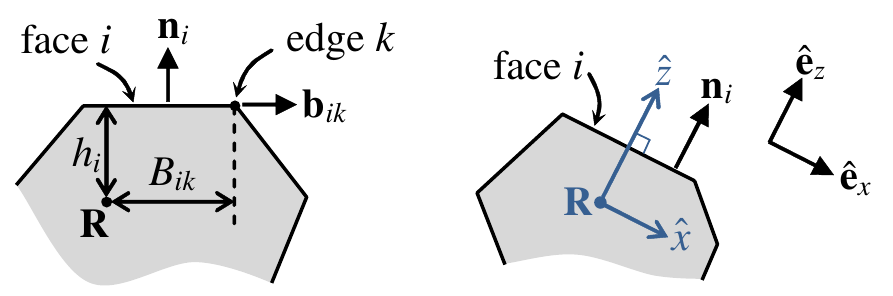}
  \caption{The meaning of the quantities $h_i$ and $B_{ik}$ defined by Eqs.~\eq{h-i} and~\eq{B-ik} (left);
  the ``tilted'' frame associated with $i$th face(right).}
  \label{fig:h-b-hat-xyz}
\end{figure}

In this Section, we will demonstrate how to simplify the functions 
$\varphi_{mnp}(\R)$, $\Phi^{(i)}_{mnp}(\R)$ and $\Omega^{(i)}_{mnp}(\R)$ 
by reducing them to the solid angles $\Omega^{(i)}(\R)$ and to some line integrals 
along the edges of the polyhedron. For this purpose, we will use two linear functions of $\R$: 
$h_i(\R)$ and $B_{ik}(\R)$, whose meaning is illustrated in Fig.~\ref{fig:h-b-hat-xyz}. 
The function $h_i(\R)$ is defined as follows:
\begin{equation} \label{h-i}
  h_i(\R) =  (\r_i - \R) \cdot \vec{n}_i \, ,
\end{equation}
where $\r_i$ is a radius vector of some point on $i$th face. (The value of $h_i$ does not 
depend on the choice of the point $\r_i$, because the vector $\vec{n}_i$ is orthogonal 
to the face.) The absolute value of $h_i$ has the meaning of the distance from $\R$
to the plane of $i$th face.

The function $B_{ik}(\R)$ has the following definition:
\begin{equation} \label{B-ik}
  B_{ik}(\R) =  (\r_k - \R) \cdot \vec{b}_{ik} \, ,
\end{equation}
where $\r_k$ is a radius vector of some point on $k$th edge. 
Since the vector $\vec{b}_{ik}$ 
is orthogonal to the edge, the choice of the point $\r_k$ does not matter. 
The absolute value of $B_{ik}$ is the distance between the projection of the point~$\R$ 
to $i$th face and the line of $k$th edge, as illustrated in Fig.~\ref{fig:h-b-hat-xyz}. 

We start from the usual recipe of reducing volume integrals to surface integrals. 
Applying Gauss's theorem to the vector
\begin{equation} \label{F-vector}
  \vec{F}(\r) =  \frac{\r-\R}{|\r-\R|} \, (x-X)^m (y-Y)^n (z-Z)^p ,
\end{equation}
one can reduce the volume integral $\phi_{mnp}$ to integrals over inclusion faces. 
The details of the derivation can be found in Appendix~\ref{app:volume-to-surface}. The answer reads:
\begin{equation} \label{phi-via-Phi}
  \phi_{mnp} = \frac{1}{m+n+p+2} \sum\limits_i h_i \, \Phi^{(i)}_{mnp} \, .
\end{equation}

Now let us consider the primitives $\Phi^{(i)}_{mnp}$ and $\Omega^{(i)}_{mnp}$ for one of the faces 
(say, $i$th face). It is convenient to choose temporarily a system of coordinates 
$\hat x, \hat y, \hat z$ related to $i$th face, so that the axis $\hat z$ is perpendicular to this face. 
For that, we choose three mutually orthogonal normal vectors 
$\hat{\vec{e}}_x, \hat{\vec{e}}_y, \hat{\vec{e}}_z$ such that $\hat{\vec{e}}_z = \vec{n}_i$, and 
define the coordinates $\hat x, \hat y, \hat z$ as follows:
\begin{subequations}
\begin{eqnarray} 
  \label{hat-xy}
  \hat x &=& (\r - \R)\cdot\hat{\vec{e}}_x ,  \qquad  \hat y = (\r - \R)\cdot\hat{\vec{e}}_y , \\
  \label{hat-z}
  \hat z &=& (\r - \R)\cdot\hat{\vec{e}}_z  \equiv  (\r - \R)\cdot\vec{n}_i  
\end{eqnarray}
\end{subequations}
(see Fig.~\ref{fig:h-b-hat-xyz}).
Then, let us define quantities $\hat\Phi^{(i)}_{mnp}$, $\hat\Omega^{(i)}_{mnp}$ and 
$\hat L^{(k)}_{mnp}$ similar to the primitives $\Phi^{(i)}_{mnp}$, $\Omega^{(i)}_{mnp}$, 
$L^{(k)}_{mnp}$ but related to the ``tilted'' axes $\hat x, \hat y, \hat z$:
\begin{subequations}
\begin{eqnarray} 
  \label{hat-Phi-i-mnp-def}
  \hat\Phi^{(i)}_{mnp} &=& \iint\limits_{\text{face} \, i} 
  \frac{\hat x^m \hat y^n \hat z^p}{\hat r} \, dS , \\
  \label{hat-Omega-i-mnp-def} 
  \hat\Omega^{(i)}_{mnp} &=& - \iint\limits_{\text{face} \, i} 
  \hat x^m \hat y^n \hat z^p \, \frac{\hat z}{\hat r^3} \, dS , \\
  \label{hat-L-k-mnp-def}
  \hat L^{(k)}_{mnp} &=& \int\limits_{\text{A}_k}^{\text{B}_k}  
  \frac{\hat x^m \hat y^n \hat z^p}{\hat r} \, dl ,
\end{eqnarray}
\end{subequations}
where $\hat r = \left( \hat x^2 + \hat y^2 + \hat z^2 \right)^{1/2} = |\r-\R|$. 
(The considered edges are those that surround $i$th face.) 
Obviously, the ``untilted'' quantities are linear combinations of ``tilted'' ones:
\begin{subequations}
\begin{eqnarray} 
  \label{hat-Phi-via-Phi}
  \Phi^{(i)}_{mnp} &=& \sum\limits_{m',n',p'} 
  T^{m'n'p'}_{mnp} \hat\Phi^{(i)}_{m'n'p'} \, , \\
  \label{hat-Omega-via-Omega}
  \Omega^{(i)}_{mnp} &=& \sum\limits_{m',n',p'} 
  T^{m'n'p'}_{mnp} \hat\Omega^{(i)}_{m'n'p'} \, ,
\end{eqnarray}
\end{subequations}
with coefficients $T^{m'n'p'}_{mnp}$ defined by the identity
\begin{equation} \label{T-identity}
  (x-X)^m (y-Y)^n (z-Z)^p = \sum\limits_{m',n',p'} 
  T^{m'n'p'}_{mnp} \hat x^{m'} \hat y^{n'} \hat z^{p'} .
\end{equation}

Note that, at $m=n=p=0$, the quantities $\Phi^{(i)}_{mnp}$, $\Omega^{(i)}_{mnp}$ and 
$L^{(k)}_{mnp}$ are invariant with respect to the choice of coordinates. Consequently,
\begin{subequations} \label{invariant-at-zero-mnp}
\begin{eqnarray} 
  \hat\Phi^{(i)}_{000} &=& \Phi^{(i)}_{000} \, , \\
  \hat\Omega^{(i)}_{000} &=& \Omega^{(i)}_{000} = \Omega^{(i)} , \\
  \hat L^{(k)}_{000} &=&  L^{(k)}_{000} =  L^{(k)} .
\end{eqnarray}
\end{subequations}

Then, it is easy to get rid of the index $p$ in $\hat\Phi^{(i)}_{mnp}$ and $\hat\Omega^{(i)}_{mnp}$. 
One can see from Eqs.~\eq{h-i} and~\eq{hat-z} that 
$i$th face lies in a plane of constant coordinate $\hat z=h_i$. One can therefore take the factor 
$\hat z^p$ outside the integral sign in Eqs.~\eq{hat-Phi-i-mnp-def}, \eq{hat-Omega-i-mnp-def}, 
what gives
\begin{equation} \label{p-to-zero}
  \hat\Phi  ^{(i)}_{mnp} = h_i^p \, \hat\Phi  ^{(i)}_{mn0} \, ,  \qquad  
  \hat\Omega^{(i)}_{mnp} = h_i^p \, \hat\Omega^{(i)}_{mn0} \, .
\end{equation}

Applying the planar variant of Gauss's theorem to the vector $(G_{\hat x},G_{\hat y})$ defined as
\begin{equation}
  G_{\hat x} (\hat x, \hat y) = \frac{\hat x^{m+1} \hat y^n}{\hat r} \, , \qquad 
  G_{\hat y} (\hat x, \hat y) = \frac{\hat x^m \hat y^{n+1}}{\hat r} \, ,
\end{equation}
one can reduce the surface integral $\hat\Phi^{(i)}_{mn0}$ 
to the quantity $\hat\Omega^{(i)}_{mn0}$ and 
line integrals $\hat L^{(k)}_{mn0}$. The details of the calculation can be found in Appendix~\ref{app:Phi-to-Omega}, 
and the following result was obtained:
\begin{equation} \label{Phi-via-Omega-L}
  \hat\Phi^{(i)}_{mn0} 
  = \frac{h_i}{m+n+1} \, \hat\Omega^{(i)}_{mn0}  
  + \sum\limits_k \frac{B_{ik}}{m+n+1} \, \hat L^{(k)}_{mn0} \, .
\end{equation}
Here and below in this section, the index $k$ runs over edges adjacent to $i$th face.

The next step is decreasing the index $n$ at $\hat\Omega^{(i)}_{mn0}$. If~$n \geq 2$, one can 
apply the following relation derived in Appendix~\ref{app:decreasing-n},
\begin{multline} \label{Omega-reduce-n}
  \hat\Omega^{(i)}_{mn0} = 
  - \hat\Omega^{(i)}_{m+2,n-2,0} 
  - \frac{m+n}{m+n-1} \, h_i^2 \, \hat\Omega^{(i)}_{m,n-2,0} \\
  - h_i \sum\limits_k \frac{B_{ik}}{m+n-1} \, \hat L^{(k)}_{m,n-2,0} \, .
\end{multline}
This recursive relation makes it possible to lower the second index at $\Omega$ down to $n=1$ or $n=0$.

In the case of $n=1$, the expression for $\hat\Omega^{(i)}_{m10}$ can be simplified 
by direct integration over $\hat y$ (see details in Appendix~\ref{app:Omega-m10}), yielding
\begin{equation} \label{Omega-m10-answer}
  \hat\Omega^{(i)}_{m10} = h_i \sum\limits_k (\vec{b}_{ik}\cdot\hat{\vec{e}}_y) \, \hat L^{(k)}_{m00} \, .
\end{equation}

The last case to consider is that of $n=0$. 
If $m \geq 2$, one can reduce $m$ by two as follows:
\begin{equation} \label{Omega-reduce-m}
  \hat\Omega^{(i)}_{m00} = 
  - h_i^2 \, \hat\Omega^{(i)}_{m-2,0,0} 
  - h_i \sum\limits_k (\vec{b}_{ik}\cdot\hat{\vec{e}}_y) \, \hat L^{(k)}_{m-2,1,0} \, .
\end{equation}
This relation is derived in Appendix~\ref{app:decreasing-m}. Applying Eq.~(\ref{Omega-reduce-m}) repeatedly, 
one arrives finally at $\hat\Omega^{(i)}_{100}$ or $\hat\Omega^{(i)}_{000} \equiv \Omega^{(i)}$ 
plus line integrals.

Finally, the value of $\hat\Omega^{(i)}_{100}$ can be found in full analogy 
with Eq.~(\ref{Omega-m10-answer}):
\begin{equation} \label{Omega-100-answer}
  \hat\Omega^{(i)}_{100} 
  = h_i \sum\limits_k (\vec{b}_{ik}\cdot\hat{\vec{e}}_x) \,  L^{(k)} \, .
\end{equation}

Thus, we have seen in this section that any of quantities $\phi_{mnp}$, $\Phi^{(i)}_{mnp}$, 
$\Omega^{(i)}_{mnp}$ can be reduced to line integrals over inclusion edges and/or 
to solid angles $\Omega^{(i)}$. A simple but not so trivial example of such reducing will be 
considered in Subsection~\ref{sec:reducing-example}. 
Then, in Section~\ref{sec:L}, we will know how to evaluate the line integrals.

\subsection{Example: evaluation of $\Omega^{(i)}_{001}$}
\label{sec:reducing-example}

\begin{figure}
  \includegraphics[width=0.7\columnwidth]{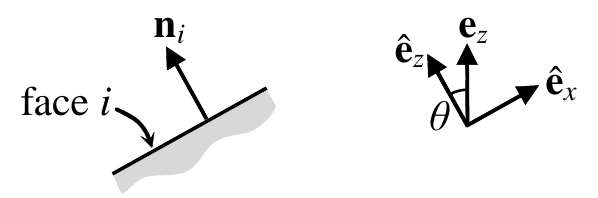}
  \caption{Unit vectors $\hat{\vec{e}}_z$ and $\hat{\vec{e}}_x$ as defined by Eqs.~\eq{example-hat-ez} and~\eq{example-hat-ex}.}
  \label{fig:theta}
\end{figure}

Here we will apply the above-formulated considerations to finding an analytical expression 
for the quantity $\Omega^{(i)}_{001}$. The first step is choosing the ``tilted'' 
frame associated with $i$th face of the polyhedron, i.~e., choosing 
three mutually orthogonal unit vectors $\hat{\vec{e}}_x$, $\hat{\vec{e}}_y$, 
$\hat{\vec{e}}_z$ along the axes of the ``tilted'' frame. 
The vector $\hat{\vec{e}}_z$ should be orthogonal to $i$th face:
\begin{equation} \label{example-hat-ez}
  \hat{\vec{e}}_z = \vec{n}_i \, .
\end{equation}
We choose the vector $\hat{\vec{e}}_x$ in the plane spanned onto $\vec{e}_z$ 
and $\hat{\vec{e}}_z$ (see Fig.~\ref{fig:theta}):
\begin{equation} \label{example-hat-ex}
  \hat{\vec{e}}_x = \frac{ \vec{e}_z - \hat{\vec{e}}_z \cos\theta }{\sin\theta} \, ,
\end{equation}
where $\theta$ is the angle between $\vec{e}_z$ and $\hat{\vec{e}}_z$:
\begin{equation} 
  \cos\theta  =  \vec{e}_z \cdot \hat{\vec{e}}_z  
  \equiv  \vec{e}_z \cdot \vec{n}_i  \equiv  n_{iz} \, .
\end{equation}
And the vector $\hat{\vec{e}}_y$ is to be orthogonal to $\hat{\vec{e}}_x$ 
and~$\hat{\vec{e}}_z$.

Then, according to Eq.~\eq{hat-Omega-via-Omega}, $\Omega^{(i)}_{001}$ is 
a linear combination of the ``tilted'' quantities $\hat\Omega^{(i)}_{100}$, 
$\hat\Omega^{(i)}_{010}$ and $\hat\Omega^{(i)}_{001}$, taken with their respective 
coefficients $T^{100}_{001}$, $T^{010}_{001}$ and $T^{001}_{001}$. In order to 
find these coefficients from Eq.~\eq{T-identity}, one should represent the value of 
$z-Z$ via the ``tilted'' coordinates $\hat x, \hat y, \hat z$:
\begin{equation} 
  z-Z = (\r-\R)\cdot\vec{e}_z 
  = (\hat x\hat{\vec{e}}_x + \hat y\hat{\vec{e}}_y + \hat z\hat{\vec{e}}_z)\cdot\vec{e}_z
  = \hat x \sin\theta + \hat z \cos\theta ;
\end{equation}
whence, $T^{100}_{001} = \sin\theta$, $T^{010}_{001} = 0$, 
and $T^{001}_{001} = \cos\theta$. Therefore
\begin{equation} \label{example-Omega-via-hat-Omega}
  \Omega^{(i)}_{001} 
  = \hat\Omega^{(i)}_{100} \sin\theta 
  + \hat\Omega^{(i)}_{001} \cos\theta .
\end{equation}

Then one can evaluate the quantity $\hat\Omega^{(i)}_{100}$ 
using Eq.~\eq{Omega-100-answer}, where scalar products $\vec{b}_{ik}\cdot\hat{\vec{e}}_x$ 
can be found from Eq.~\eq{example-hat-ex} taking into account that 
the vector $\vec{b}_{ik}$ is orthogonal to $\hat{\vec{e}}_z\equiv\vec{n}_i$:
\begin{equation} 
  \vec{b}_{ik}\cdot\hat{\vec{e}}_x = \frac{\vec{b}_{ik}\cdot\vec{e}_z}{\sin\theta} 
  \equiv \frac{b_{ikz}}{\sin\theta} \, .
\end{equation}
Consequently, Eq.~\eq{Omega-100-answer} takes the following form:
\begin{equation} \label{example-hat-Omega-100}
  \hat\Omega^{(i)}_{100} 
  = h_i \sum\limits_k \frac{b_{ikz}}{\sin\theta} \, L^{(k)} .
\end{equation}

And the quantity $\hat\Omega^{(i)}_{001}$ is calculated according to Eq.~\eq{p-to-zero}:
\begin{equation} \label{example-hat-Omega-001}
  \hat\Omega^{(i)}_{001} = h_i \, \hat\Omega^{(i)}_{000} 
  \equiv h_i \, \Omega^{(i)} .
\end{equation}
Finally, collecting Eqs.~\eq{example-Omega-via-hat-Omega}, \eq{example-hat-Omega-100}, 
\eq{example-hat-Omega-001}, one can get the following answer:
\begin{equation} \label{example-answer}
  \Omega^{(i)}_{001} 
  = h_i \sum\limits_k b_{ikz} \, L^{(k)} 
  + h_i \, n_{iz} \, \Omega^{(i)} .
\end{equation}
This is an analytical representation of $\Omega^{(i)}_{001}$ in a closed form, 
because, as mentioned 
in Section~\ref{sec:primitives}, the primitives $L^{(k)}$ and $\Omega^{(i)}$ 
can be expressed through elementary functions.

\section{Evaluation of line integrals $L^{(k)}_{mnp}$}
\label{sec:L}

Let $\mathrm{A}_k$ and $\mathrm{B}_k$ be two end points of $k$th edge 
(we denote their radius vectors as $\r_A^{(k)}$ and $\r_B^{(k)}$), 
and $\vec{l}_k$ be the unit vector directed from $\mathrm{A}_k$ to $\mathrm{B}_k$. 
To find analytical expressions for the line integrals $L^{(k)}_{mnp}$, 
it is convenient to reduce them to more simple integrals $\LL^{(k)}_t$:
\begin{equation} \label{LL}
  \LL^{(k)}_t = \int\limits_{\xi_{k1}}^{\xi_{k2}} \frac{\xi^t}{\sqrt{\rho_k^2+\xi^2}} \, d\xi , 
\end{equation}
where $t$ is a non-negative integer number, $\xi = \vec{l}_k\cdot(\r-\R)$ is a coordinate 
along $k$th edge, $\rho_k$ is the distance from the point $\R$ 
to the line of $k$th edge:
\begin{equation}
  \rho_k^2 = \big|\r_A^{(k)}-\R\big|^2 - \left[ \vec{l}_k \cdot \big(\r_A^{(k)}-\R\big) \right]^2 ,
\end{equation}
$\xi_{k1}$ and $\xi_{k2}$ are the values of $\xi$ at $\mathrm{A}_k$ and $\mathrm{B}_k$:
\begin{equation} \label{xi-k12}
  \xi_{k1} = \vec{l}_k \cdot \big(\r_A^{(k)}-\R\big) ,  \qquad
  \xi_{k2} = \vec{l}_k \cdot \big(\r_B^{(k)}-\R\big) .
\end{equation}

In order to perform such reduction, let us express the position vector $\r$ 
of a point \emph{on the edge} through the coordinate $\xi$:
\begin{equation} \label{r-via-xi}
  \r(\xi) = \r^{(k)}_0 + \vec{l}_k \xi ,
\end{equation}
where the point $\r^{(k)}_0 = \left( x^{(k)}_0, y^{(k)}_0, z^{(k)}_0 \right)$ 
is the projection of the point $\R$ to $k$th edge:
\begin{equation} \label{r0}
  \r^{(k)}_0 = \r_A^{(k)} - \mathbf{l}_k \left[ \mathbf{l}_k \cdot \big(\r_A^{(k)}-\R\big) \right] .
\end{equation}
Using Eq.~(\ref{r-via-xi}), one can represent the factor $(x-X)^m (y-Y)^n (z-Z)^p$ 
that appears in the definition of $L^{(k)}_{mnp}$ as follows:
\begin{multline} \label{f-mnp-xi-1}
  (x-X)^m (y-Y)^n (z-Z)^p = \\
  \left( x^{(k)}_0 \!\! - \!\! X \!\! + l_{k,x}\xi \right)^m  
  \left( y^{(k)}_0 \!\! - \!\! Y \!\! + l_{k,y}\xi \right)^n  
  \left( z^{(k)}_0 \!\! - \!\! Z \!\! + l_{k,z}\xi \right)^p .
\end{multline}
The right hand side of Eq.~\eq{f-mnp-xi-1} is some polynomial on $\xi$:
\begin{multline} \label{f-mnp-xi-2}
  \left( x^{(k)}_0 \!\! - \!\! X \!\! + l_{k,x}\xi \right)^m  
  \left( y^{(k)}_0 \!\! - \!\! Y \!\! + l_{k,y}\xi \right)^n  
  \left( z^{(k)}_0 \!\! - \!\! Z \!\! + l_{k,z}\xi \right)^p \\
  = \sum\limits_{t=0}^{m+n+p} c_{mnp,t}(\R) \; \xi^t ,
\end{multline}
where each coefficient $c_{mnp,t}(\R)$ is a polynomial of $X,Y,Z$ of degree $m+n+p-t$. 
Substituting this series decomposition into the line integral $L^{(k)}_{mnp}$, Eq.~\eq{L-k-mnp-def}, 
and taking into account that $|\r-\R| = \sqrt{\rho_k^2+\xi^2}$ on the edge, one can easily see that
\begin{equation} \label{L-via-mathcal-L}
  L^{(k)}_{mnp} 
  = \sum\limits_{t=0}^{m+n+p} c_{mnp,t}(\R) \; \LL^{(k)}_t \, .
\end{equation}

A similar representation takes place for the ``tilted'' line integrals $\hat L^{(k)}_{mnp}$ 
defined by Eq.~\eq{hat-L-k-mnp-def}:
\begin{equation} \label{hat-L-via-mathcal-L}
  \hat L^{(k)}_{mnp} 
  = \sum\limits_{t=0}^{m+n+p} \hat c_{mnp,t}(\R) \; \LL^{(k)}_t \, .
\end{equation}
The coefficients $\hat c_{mnp,t}(\R)$ are to be obtained from the identity
\begin{multline} \label{f-mnp-xi-2-hat}
  \left( \hat x^{(k)}_0 + \hat l_{k,x} \xi \right)^m  
  \left( \hat y^{(k)}_0 + \hat l_{k,y} \xi \right)^n  
  \left( \hat z^{(k)}_0 + \hat l_{k,z} \xi \right)^p \\
  = \sum\limits_{t=0}^{m+n+p} \hat c_{mnp,t}(\R) \; \xi^t ,
\end{multline}
where
\begin{equation}
  \hat x^{(k)}_0 = \big( \r^{(k)}_0 - \R \big) \cdot \hat{\vec{e}}_x \, , \qquad
  \hat l_{k,x} = \vec{l}_k \cdot \hat{\vec{e}}_x \, ,
\end{equation}
and the same for $y$- and $z$-components.

The last question is how to find analytical expressions 
for the integrals $\LL^{(k)}_0, \LL^{(k)}_1 , \ldots$ 
As explained in Appendix~\ref{app:LL}, integration of Eq.~\eq{LL} by parts 
provides the following recursive formula for $t>1$:
\begin{multline} \label{L-n-answer}
  \LL^{(k)}_t = - \frac{t-1}{t} \, \rho_k^2 \, \LL^{(k)}_{t-2} \\
  + \frac1t \left[ \vec{l}_k \cdot \big(\r_B^{(k)}-\R\big) \right]^{t-1} \left|\r_B^{(k)}-\R\right| \\
  - \frac1t \left[ \vec{l}_k \cdot \big(\r_A^{(k)}-\R\big) \right]^{t-1} \left|\r_A^{(k)}-\R\right| ,
\end{multline}
and a simple answer for $t=1$:
\begin{equation} \label{L-1-answer}
  \LL^{(k)}_1  =  \left| \r_B^{(k)}-\R \right|  -  \left| \r_A^{(k)}-\R \right| .
\end{equation}

Eqs.~(\ref{L-n-answer}) and~(\ref{L-1-answer}) provide a possibility to evaluate 
the quantities $\LL^{(k)}_t$ for integer $t \geq 0$ in terms of 
$\LL^{(k)}_0$ and coordinates of the ends $\mathrm{A}_k$ and $\mathrm{B}_k$ of the edge. 
In its turn, $\LL^{(k)}_0$ is the same as $L^{(k)}$ 
and therefore has the analytical representation according to Eq.~(\ref{L-000-answer}). 
So, the problem of evaluating of line integrals along inclusion edges is solved.

\section{The main result}
\label{sec:main}

Here we summarize the results of Sections~\ref{sec:analogy}--\ref{sec:reducing} and present 
the analytical expression for the strain distribution that solves the problem posed 
in the Introduction.

At first, let us recall the steps carried out in the above text. In Section~\ref{sec:analogy}, 
the strain tensor $\ve_{\alpha\beta}(\R)$ was expressed in terms of second derivatives 
of the electrostatic potential $\phi(\R)$. In Section~\ref{sec:monoms}, these second derivatives, 
as well as the first derivatives and the potential itself, were represented as combinations of 
the quantities $\phi_{mnp}$, $\phi^{(mnp)}_{,\alpha}$, $\phi^{(mnp)}_{,\alpha\beta}$, which 
correspond to the charge distribution $(x-X)^m (y-Y)^n (z-Z)^p \, \chi(\r)$. These quantities, 
in their turn, were reduced in Section~\ref{sec:primitives} to four kinds of ``primitives'': 
volume integrals $\varphi_{mnp}$, surface integrals $\Phi^{(i)}_{mnp}$ and $\Omega^{(i)}_{mnp}$, 
and line integrals $L^{(k)}_{mnp}$. In Section~\ref{sec:reducing}, the volume and surface 
integrals were expressed through the solid angles $\Omega^{(i)}$ and line integrals 
$\hat L^{(k)}_{mnp}$. Finally, in Section\ref{sec:L} all the line integrals $L^{(k)}_{mnp}$ 
and $\hat L^{(k)}_{mnp}$ were reduced to the simplest line integrals $L^{(k)}$, whose analytical 
representations are given by Eq.~\eq{L-000-answer}, and to the distances from 
the point $\R$ (where the strain tensor is to be found) to vertices of the polyhedron.

Each of these steps represents itself a linear transformation whose coefficients are either 
constants (such as components of unit vectors $\vec{n}_i$ in Eq.~\eq{phi-alpha-via-phi-Phi}) 
or polynomials of $\R$ (for example, the factor $h_i=(\r_i-\R)\cdot\vec{n}_i$ 
in Eq.~\eq{phi-via-Phi}). 

As a result, the strain tensor $\ve_{\alpha\beta}(\R)$ appears to be a linear combination of 
three kinds of analytical functions:
\begin{itemize}
\item surface integrals (potentials of the inclusion faces covered by uniform double layers) 
$\Omega^{(i)}(\R) \equiv \Omega^{(i)}_{000}(\R)$ defined according 
to Eq.~\eq{Omega-i-mnp-def} and having a geometrical meaning of solid angles;
\item line integrals (potentials of uniformly charged edges of the inclusion) 
$L^{(k)}(\R) \equiv L^{(k)}_{000}(\R)$ that have the analytical form according to Eq.~\eq{L-000-answer};
\item distances $|\R-\r_s|$ between the point $\R$ and vertices $\r_s$ of the inclusion 
(we will use the index $s$ to number the vertices).
\end{itemize}
The same is true for the potential $\phi(\R)$ and its first and second derivatives on $\R$. 
The coefficients at quantities $\Omega^{(i)}$, $L^{(k)}$ and $|\R-\r_s|$ come from substituting 
polynomial expressions one into the other, and therefore should be polynomials of $\R$.

The facts described in this Section constitute the main result of the present study. 
In Subsection~\ref{sec:main-expression} we represent them in a more formal way, writing down 
the analytical expressions that define the distributions of the strain, the potential, 
and derivatives of the potential, up to some polynomial coefficients. 
Then, in Subsection~\ref{sec:main-algorithm} we will show how to find these polynomial coefficients.

\subsection{The main result: Analytical expressions for the strain distribution, the~potential, and its derivatives}
\label{sec:main-expression}

Based on the above-presented considerations, one can formulate a universal expression 
for the strain distribution $\ve_{\alpha\beta}(\R)$ induced by a polyhedral inclusion with 
a lattice misfit $\ve_0(\R)$ that depends on coordinates polynomially:
\begin{multline} \label{main-result}
  \ve_{\alpha\beta}(\R) = 
       \sum\limits_{\substack{i \\ \text{(faces)}}}    \mathcal{A}^{(i)}_{\ve\alpha\beta} \, \Omega^{(i)} 
  \; + \sum\limits_{\substack{k \\ \text{(edges)}}}    \mathcal{B}^{(k)}_{\ve\alpha\beta} \, L^{(k)}          \\
  \; + \sum\limits_{\substack{s \\ \text{(vertices)}}} \mathcal{C}^{(s)}_{\ve\alpha\beta} \, |\R-\r_{s}| .
\end{multline}
Here summations take place over all faces (index $i$), edges (index $k$), 
and vertices (index $s$) of the polyhedron. 
$\Omega^{(i)}(\R)$ and $L^{(k)}(\R)$ are analytical functions described in 
Section~\ref{sec:primitives}: $\Omega^{(i)}$ has a geometrical meaning of a solid angle, 
and $L^{(k)}$ is defined by Eq.~\eq{L-000-answer}. 
$\r_s$ is a position vector of $s$th vertex. 
$\mathcal{A}^{(i)}_{\ve\alpha\beta}(\R)$, $\mathcal{B}^{(k)}_{\ve\alpha\beta}(\R)$, 
$\mathcal{C}^{(s)}_{\ve\alpha\beta}(\R)$ are some polynomial functions of $\R$ that 
are defined by orientations of faces and edges and by coordinates of vertices. 
The degrees of polynomials $\mathcal{A}^{(i)}_{\ve\alpha\beta}$ 
and $\mathcal{B}^{(k)}_{\ve\alpha\beta}$ are equal to $N$, i.~e., to the degree of the 
polynomial $f(\r)$ that determines the distribution of lattice misfit, 
whereas degrees of $\mathcal{C}^{(s)}_{\ve\alpha\beta}$ are equal to $N-1$. 
(if $N=0$, that corresponds to a constant lattice misfit, 
then $\mathcal{C}^{(s)}_{\ve\alpha\beta}=0$ at any $\R$.) 
A recipe for calculation of the polynomial coefficients of 
$\mathcal{A}^{(i)}_{\ve\alpha\beta}$, $\mathcal{B}^{(k)}_{\ve\alpha\beta}$ 
and $\mathcal{C}^{(s)}_{\ve\alpha\beta}$ will be described 
in Subsection~\ref{sec:main-algorithm}. Of course, Eq.~\eq{main-result} is valid 
only if the assumptions listed in the Introduction are fulfilled.

In a fully analogous manner, one can write down a similar expression for 
the electrostatic potential $\phi$ [a solution of Poisson's equation~\eq{phi-Poisson}] 
produced by a charged polyhedron with a non-uniform charge distribution $\ve(\r)$, 
which is a polynomial function of coordinates:
\begin{multline} \label{main-result-phi}
  \phi(\R) = 
       \sum\limits_{\substack{i \\ \text{(faces)}}}    \mathcal{A}^{(i)}_{\phi} \, \Omega^{(i)} 
  \; + \sum\limits_{\substack{k \\ \text{(edges)}}}    \mathcal{B}^{(k)}_{\phi} \, L^{(k)}          \\
  \; + \sum\limits_{\substack{s \\ \text{(vertices)}}} \mathcal{C}^{(s)}_{\phi} \, |\R-\r_{s}| ,
\end{multline}
where polynomials $\mathcal{A}^{(i)}_{\phi}(\R)$ and $\mathcal{B}^{(k)}_{\phi}(\R)$ 
have the degree $N+2$, and polynomials $\mathcal{C}^{(s)}_{\phi}(\R)$ 
have the degree $N+1$.

The first and second derivatives of the potential $\phi(\R)$ also have similar 
representations:
\begin{multline} \label{main-result-phi-deriv}
  \frac{\d\phi(\R)}{\d R_\alpha} = 
       \sum\limits_{\substack{i \\ \text{(faces)}}}    \mathcal{A}^{(i)}_{\phi\alpha} \, \Omega^{(i)} 
  \; + \sum\limits_{\substack{k \\ \text{(edges)}}}    \mathcal{B}^{(k)}_{\phi\alpha} \, L^{(k)}          \\
  \; + \sum\limits_{\substack{s \\ \text{(vertices)}}} \mathcal{C}^{(s)}_{\phi\alpha} \, |\R-\r_{s}| 
\end{multline}
and
\begin{multline} \label{main-result-phi-2deriv}
  \frac{\d^2\phi(\R)}{\d R_\alpha \d R_\beta} = 
       \sum\limits_{\substack{i \\ \text{(faces)}}}    \mathcal{A}^{(i)}_{\phi\alpha\beta} \, \Omega^{(i)} 
  \; + \sum\limits_{\substack{k \\ \text{(edges)}}}    \mathcal{B}^{(k)}_{\phi\alpha\beta} \, L^{(k)}          \\
  \; + \sum\limits_{\substack{s \\ \text{(vertices)}}} \mathcal{C}^{(s)}_{\phi\alpha\beta} \, |\R-\r_{s}| ,
\end{multline}
where $\mathcal{A}^{(i)}_{\phi\alpha}(\R)$ and $\mathcal{B}^{(k)}_{\phi\alpha}(\R)$ 
are polynomials of degree $N+1$; 
$\mathcal{C}^{(s)}_{\phi\alpha}(\R)$, $\mathcal{A}^{(i)}_{\phi\alpha\beta}(\R)$ 
and $\mathcal{B}^{(k)}_{\phi\alpha\beta}(\R)$ are polynomials of degree $N$; 
and $\mathcal{C}^{(s)}_{\phi\alpha\beta}(\R)$ are polynomials of degree $N-1$. 
The coefficients of all these polynomials are to be calculated by the method of 
Subsection~\ref{sec:main-algorithm}.


\subsection{The main result: Algorithm of finding the polynomial coefficients}
\label{sec:main-algorithm}

Equations~\eq{main-result}--\eq{main-result-phi-2deriv} determine the sought-for 
functions $\ve_{\alpha\beta}(\R)$, $\phi(\R)$, $\d\phi(\R)/\d R_\alpha$ 
and $\d^2\phi(\R)/\d R_\alpha \d R_\beta$ up to the coefficients of the corresponding 
polynomials $\mathcal{A}$, $\mathcal{B}$, $\mathcal{C}$. Here we represent the 
algorithm (a sequence of steps) for calculating these coefficients. The algorithm 
is just a summary of equations obtained in previous sections.

On the input of the algorithm there are: 
the polynomial function $f(x,y,z)$, which defines the distribution of eigenstrain 
inside the inclusion, 
and the geometry of the inclusion---unit vectors 
$\vec{n}_i$, $\vec{b}_{ik}$, $\vec{l}_k$ and coordinates of vertices.

On the output there are polynomials $\mathcal{A}^{(i)}_{\ve\alpha\beta}(\R)$, 
$\mathcal{B}^{(k)}_{\ve\alpha\beta}(\R)$, $\mathcal{C}^{(s)}_{\ve\alpha\beta}(\R)$ 
that determine the distribution of elastic strain $\ve_{\alpha\beta}(\R)$ 
according to Eq.~(\ref{main-result}). For finding the polynomials related to 
the potential and its derivatives, one should modify Steps~1--3, 
as explained in the end of this Subsection.

The algorithm consists of eleven steps listed below. 
Each step contains only arithmetic operations on polynomials. 
Therefore, it is easy to implement the algorithm in a program code.

Step 1. Express strain components $\ve_{\alpha\beta}$ via second derivatives 
$\d^2\phi / \d R_\alpha \d R_\beta$ using Eq.~(\ref{epsilon-via-potential}); 
express the function $\ve_0$ that appears in the right-hand side of 
Eq.~(\ref{epsilon-via-potential}) via solid angles $\Omega^{(i)}$ 
using Eq.~\eq{rho-via-Omega}.

Step 2. Find the coefficients $\tilde C_{mnp}(\R)$ from Eq.~(\ref{tilde-C}), 
and represent the second derivatives $\d^2\phi / \d R_\alpha \d R_\beta$ 
as combinations of $\phi^{(mnp)}_{,\alpha\beta}$ using Eq.~(\ref{phi-2deriv-via-phi-mnp}).

Step 3. Express each of the quantities $\phi^{(mnp)}_{,\alpha\beta}$ through integrals 
$\phi_{m'n'p'}$, $\Phi^{(i)}_{m'n'p'}$, $\Omega^{(i)}_{m'n'p'}$ and $L^{(k)}_{m'n'p'}$ 
with different $m',n',p'$ according to Eqs.~(\ref{phi-alpha-beta-via-primitives}), 
\eq{phi-mnp-low2-calc}, \eq{Phi-mnp-low-calc}.

Step 4. Reduce each of the volume integrals $\phi_{mnp}$ to a combination of surface 
integrals $\Phi^{(i)}_{mnp}$ using Eq.~(\ref{phi-via-Phi}).
(After this step the strain tensor, as a function of coordinates, is represented 
as a linear combination of functions $\Phi^{(i)}_{mnp}$, $\Omega^{(i)}_{mnp}$, 
$L^{(k)}_{mnp}$ with polynomial coefficients.)

\emph{Steps 5--9 should be performed for each face $i$ of the inclusion:}

Step 5. Choose a ``tilted'' coordinate system $\hat x, \hat y, \hat z$ with axis $\hat z$ 
along the normal $\vec{n}_i$ to $i$th face. Then, 
transform the integrals $\Phi^{(i)}_{mnp}$ and $\Omega^{(i)}_{mnp}$ into their ``tilted'' versions 
$\hat\Phi^{(i)}_{mnp}$ and $\hat\Omega^{(i)}_{mnp}$ using Eqs.~(\ref{hat-Phi-via-Phi}) 
and~(\ref{hat-Omega-via-Omega}). The coefficients of transformation $T^{m'n'p'}_{mnp}$ 
are to be found from the identity~\eq{T-identity}.

Step 6. Decrease the index $p$ at $\hat\Phi^{(i)}_{mnp}$ and $\hat\Omega^{(i)}_{mnp}$ 
down to zero by means of Eq.~(\ref{p-to-zero}).

Step 7. Express the integrals $\hat\Phi^{(i)}_{mn0}$ via $\hat\Omega^{(i)}_{mn0}$ 
and $\hat L^{(k)}_{mn0}$ using Eq.~(\ref{Phi-via-Omega-L}). 
(After this step, the strain distribution is expressed in terms of surface integrals 
$\hat\Omega^{(i)}_{mn0}$ and line integrals $L^{(k)}_{mnp}$, $\hat L^{(k)}_{mn0}$.)

Step 8. For each $\hat\Omega^{(i)}_{mn0}$, decrease the index $n$ recursively 
down to $n=1$ or $n=0$ by means of Eq.~(\ref{Omega-reduce-n}). 
Then, if $n=1$, express the quantity $\hat\Omega^{(i)}_{m10}$ via $\hat L^{(k)}_{m00}$ 
using Eq.~(\ref{Omega-m10-answer}).

Step 9. For each $\hat\Omega^{(i)}_{m00}$, decrease the index $m$ recursively 
down to $m=1$ or $m=0$ by means of Eq.~(\ref{Omega-reduce-m}). 
Then, if $m=1$, express the quantity $\hat\Omega^{(i)}_{100}$ via $L^{(k)}$ 
using Eq.~(\ref{Omega-100-answer}). 
(After this step, the strain distribution appears to be a combination of 
solid angles $\Omega^{(i)}$ and line integrals $L^{(k)}_{mnp}$, $\hat L^{(k)}_{mn0}$.)

Step 10. For each edge $k$ of the inclusion, decompose each line integral $L^{(k)}_{mnp}$ 
into a series $c_{mnp,0} \mathcal{L}^{(k)}_0 + c_{mnp,1} \mathcal{L}^{(k)}_1 + \ldots$ 
according to Eq.~\eq{L-via-mathcal-L}, where coefficients 
$c_{mnp,0}(\R), c_{mnp,1}(\R), \ldots$ are to be found 
from Eq.~(\ref{f-mnp-xi-2}). 
Do the same with ``tilted'' line integrals $\hat L^{(k)}_{mn0}$ 
by means of Eqs.~\eq{hat-L-via-mathcal-L}, \eq{f-mnp-xi-2-hat}.

Step 11. For each edge $k$ and each line integral $\mathcal{L}^{(k)}_t$, 
decrease the index $t$ recursively down to $t=1$ or $t=0$ by means of Eq.~(\ref{L-n-answer}). 
Then, if $t=1$, evaluate the integral $\mathcal{L}^{(k)}_1$ using Eq.~(\ref{L-1-answer}).

After performing Steps 1--11, one obtains the strain distribution 
$\ve_{\alpha\beta}(\R)$ in the form of a combination of solid angles 
$\Omega^{(i)}(\R)$, logarithmic expressions $L^{(k)}(\R)$, and distances from $\R$ 
to inclusion vertices, i.~e., in the form of 
Eq.~(\ref{main-result}) with the \emph{known} polynomial coefficients 
$\mathcal{A}^{(i)}_{\ve\alpha\beta}(\R)$, $\mathcal{B}^{(k)}_{\ve\alpha\beta}(\R)$, 
$\mathcal{C}^{(s)}_{\ve\alpha\beta}(\R)$. 
Hence, these steps actually constitute an algorithm for calculating 
the polynomials $\mathcal{A}$, $\mathcal{B}$, $\mathcal{C}$.

In Appendix~\ref{sec:pseudocode}, one can find an example of implementation of this algorithm in the form of an informal ``pseudocode''.

In a slightly modified form, this algorithm can be also used to calculate 
the potential $\phi(\R)$, its derivatives and its second derivatives. For that, 
it is enough to omit Step~1 and make obvious modifications of Steps~2 and~3. 
Namely, for finding an expression for the potential $\phi(\R)$, one can use 
Eq.~\eq{phi-via-phi-mnp} instead of Eq.~\eq{phi-2deriv-via-phi-mnp}, and omit Step~3. 
For finding the expressions for the first derivatives $\d\phi(\R)/\d R_\alpha$, 
one can use Eq.~\eq{phi-deriv-via-phi-mnp} instead of Eq.~\eq{phi-2deriv-via-phi-mnp} 
at Step~2, 
and Eqs.~\eq{phi-alpha-via-phi-Phi}, \eq{phi-mnp-low-calc} 
instead of Eqs.~\eq{phi-alpha-beta-via-primitives}, \eq{phi-mnp-low2-calc}, \eq{Phi-mnp-low-calc} at Step~3.

\section{Examples}
\label{sec:examples}

This Section has two aims: (i) to illustrate the results of Section~\ref{sec:main} 
on concrete examples, and (ii) to provide the evidence that our results are correct.

\subsection{Constant lattice misfit}
\label{sec:examples-homogeneous}

For the simplest example, let us suppose that the lattice parameter is the same in 
any point of the inclusion. Therefore
\begin{equation}
  f = \text{const} ,
\end{equation}
and the lattice misfit $\ve_0$ is equal to $f$ inside the inclusion, 
and to 0 outside. Applying the algorithm of Section~\ref{sec:main} to this case, 
one can see that the answer for the strain distribution in the form~(\ref{main-result}) 
appears already after Step~3:
\begin{multline} \label{uniform-eps-full}
  \ve_{\alpha\beta}(\R)  
  =  - \Lambda f \sum\limits_i n_{i\alpha} n_{i\beta} \Omega^{(i)}(\R) \\
  - \Lambda f \sum\limits_k \lambda^{(k)}_{\alpha\beta} L^{(k)}(\R) 
  + \frac{f}{4\pi} \, \delta_{\alpha\beta} \sum\limits_i \Omega^{(i)}(\R) ,
\end{multline}
where values of $\Lambda$ and $\lambda^{(k)}_{\alpha\beta}$ are defined by 
Eqs.~\eq{Lambda-via-nu} and~\eq{lambda-k-alpha-beta}. 
A similar result was reported in Ref.~\onlinecite{Nenashev2010}.

A comparison between Eq.~\eq{main-result} and Eq.~\eq{uniform-eps-full} provides 
the values of the coefficients $\mathcal{A}$, $\mathcal{B}$, $\mathcal{C}$ 
for the case of a constant lattice misfit inside the inclusion:
\begin{subequations} \label{uniform-eps}
\begin{align}
  \mathcal{A}^{(i)}_{\ve\alpha\beta}  &=  - \Lambda f n_{i\alpha} n_{i\beta} + \delta_{\alpha\beta} f/4\pi , \\
  \label{uniform-eps-B}
  \mathcal{B}^{(k)}_{\ve\alpha\beta}  &=  - \Lambda f \lambda^{(k)}_{\alpha\beta} \, , \\
  \mathcal{C}^{(s)}_{\ve\alpha\beta}  &=  0 \, .
\end{align}
\end{subequations}

\begin{figure}
  \includegraphics[width=\columnwidth]{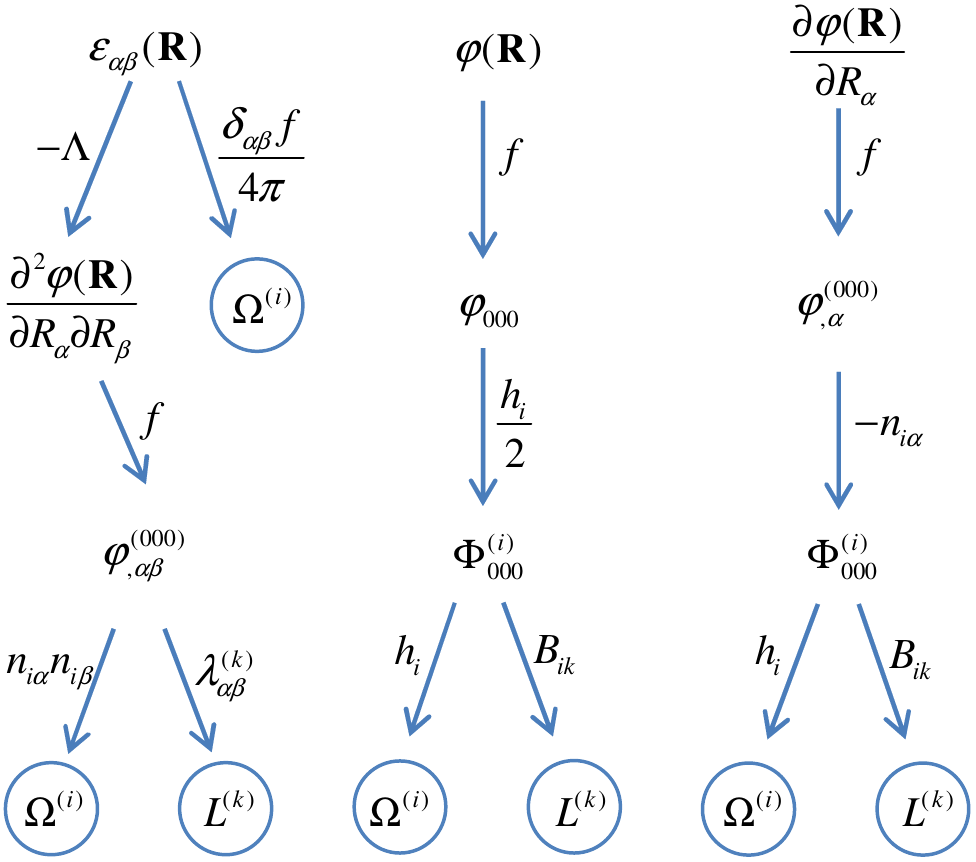}
  \caption{Schematic representation of the algorithm of calculating 
  the strain tensor $\ve_{\alpha\beta}$ (left part), 
  the potential $\phi$ (middle part), 
  derivatives of the potential $\d\phi/\d R_\alpha$ (right part), 
  and second derivatives of the potential $\d^2\phi/\d R_\alpha\d R_\beta$ 
  (a subgraph of the left part) 
  in the case of a uniform eigenstrain: $f=\text{const}$. 
  See details in Subsection~\ref{sec:examples-homogeneous}.}
  \label{fig:graph-constant-f}
\end{figure}

For better understanding of the way leading to this answer, one can visualize 
the algorithm in the form of a tree, as shown in the left part 
of Fig.~\ref{fig:graph-constant-f}. The sought-for function $\ve_{\alpha\beta}$ 
is placed at the top of this graph. 
At Step~1, this function acquires the following representation:
\begin{equation} \label{uniform-step1}
  \ve_{\alpha\beta} = -\Lambda \; \frac{\d^2\phi}{\d R_\alpha \d R_\beta} 
  + \sum\limits_i \frac{\delta_{\alpha\beta}f}{4\pi} \; \Omega^{(i)} ,
\end{equation}
that is symbolized in the graph by arrows connecting $\ve_{\alpha\beta}$ 
to $\d^2\phi/\d R_\alpha \d R_\beta$ and to $\Omega^{(i)}$. A value near an arrow 
means the factor, with which the expression on the head of the arrow contributes to 
the expression on the tail. Then, Step~2 represents $\d^2\phi/\d R_\alpha \d R_\beta$ 
as follows:
\begin{equation} \label{uniform-step2}
  \frac{\d^2\phi}{\d R_\alpha \d R_\beta} = f \; \phi^{(000)}_{,\alpha\beta} .
\end{equation}
This connection is shown in the graph by the arrow with mark ``$f$''. 
Finally, Step~3 provides the expression for $\phi^{(000)}_{,\alpha\beta}$:
\begin{equation} \label{uniform-step3}
  \phi^{(000)}_{,\alpha\beta} = \sum\limits_i n_{i\alpha} n_{i\beta} \; \Omega^{(i)} 
  + \sum\limits_l \lambda^{(k)}_{\alpha\beta} \; L^{(k)} ,
\end{equation}
which is depicted by two lower arrows in the graph.

The formulas for the coefficients $\mathcal{A}$, $\mathcal{B}$, $\mathcal{C}$ 
can be easily constructed on the basis of this graph. 
Each appearance of $\Omega^{(i)}$ or $L^{(k)}$ 
in the graph provides the contribution to $\mathcal{A}^{(i)}_{\ve\alpha\beta}$ 
or $\mathcal{B}^{(k)}_{\ve\alpha\beta}$, correspondingly. The value of the contribution 
is equal to the product of the quantities on arrows along the way from the root 
of the tree to $\Omega^{(i)}$ or $L^{(k)}$. For example, the way to $L^{(k)}$ 
consists of three arrows with values $-\Lambda$, $f$ and $\lambda^{(k)}_{\alpha\beta}$ 
on them; hence, their product $-\Lambda f \lambda^{(k)}_{\alpha\beta}$ appears 
in Eq.~\eq{uniform-eps-B}.

In a similar manner, one can calculate the potential of a uniformly charged 
polyhedral body $\phi(\R)$, and its first and second derivatives. 
Since $f$ is a constant ($N=0$), only primitives with $m=n=p=0$ will appear 
in the course of the algorithm. 
For this reason, one can omit Steps 5, 6, 8--11 (as well as Step~1, 
which is needed only for calculating the strain). 
The result for the potential is
\begin{subequations} \label{uniform-fi}
\begin{align}
  \mathcal{A}^{(i)}_{\phi}(\R)  &=  \frac{f}{2} (h_i)^2 , \\
  \mathcal{B}^{(k)}_{\phi}(\R)  &=  \frac{f}{2} \big( h_{i_1} B_{i_1 k} + h_{i_2} B_{i_2 k} \big) , \\
  \mathcal{C}^{(s)}_{\phi}(\R)  &=  0 \, ,
\end{align}
\end{subequations}
where $i_1$ and $i_2$ are numbers of the two faces that intersect at $k$th edge; 
the linear functions $h_i(\R)$ and $B_{ik}(\R)$ are defined by Eqs.~\eq{h-i} and~\eq{B-ik}, 
correspondingly. Expressions~\eq{uniform-fi} are to be inserted into the general formula~\eq{main-result-phi}.

For the first derivatives of the potential, the answer is:
\begin{subequations} \label{uniform-fi-deriv}
\begin{align}
  \mathcal{A}^{(i)}_{\phi\alpha}(\R)  &=  -f n_{i\alpha} h_i \, , \\
  \mathcal{B}^{(k)}_{\phi\alpha}(\R)  &=  -f \big( n_{i_1\alpha} B_{i_1 k} + n_{i_2\alpha} B_{i_2 k} \big) \, , \\
  \mathcal{C}^{(s)}_{\phi\alpha}(\R)  &=  0 \, ,
\end{align}
\end{subequations}
and there is the answer for the second derivatives of the potential:
\begin{subequations} \label{uniform-fi-2deriv}
\begin{align}
  \mathcal{A}^{(i)}_{\phi\alpha\beta}  &=  f n_{i\alpha} n_{i\beta} \, , \\
  \mathcal{B}^{(k)}_{\phi\alpha\beta}  &=  f \lambda^{(k)}_{\alpha\beta} \, , \\
  \mathcal{C}^{(s)}_{\phi\alpha\beta}  &=  0 \, .
\end{align}
\end{subequations}

In Fig.~\ref{fig:graph-constant-f} one can see also graphical representations 
for the potential $\phi$ 
(middle part), its derivative $\d\phi/\d R_\alpha$ (right part), 
and its second derivative $\d^2\phi/\d R_\alpha\d R_\beta$ (the part of 
the left graph, beginning from $\d^2\phi/\d R_\alpha\d R_\beta$).

Our expressions~\eq{uniform-fi}--\eq{uniform-fi-2deriv} for the potential of a homogeneous polyhedron and its first and second derivatives 
are the same as the results by Werner and Scheeres.\cite{Werner1996} 
One can consider the coincidence of our results for $f=\text{const}$ 
with the published ones as a simplest test on the correctness of our approach. 
According to the present Subsection, this test is successfully passed.   

Also we have performed a comparison with analytical calculations made by Glas\cite{Glas2001} 
for the case of a truncated pyramid with a constant misfit strain. 
The results of calculations shown in Fig.~3b of Ref.~\onlinecite{Glas2001} are fully reproduced by our method. 
For more details, see Appendix~\ref{sec:Glas}.

\subsection{Constant, vertically directed gradient of the lattice misfit}
\label{sec:examples-linear-z}

\begin{figure}
  \includegraphics[width=\columnwidth]{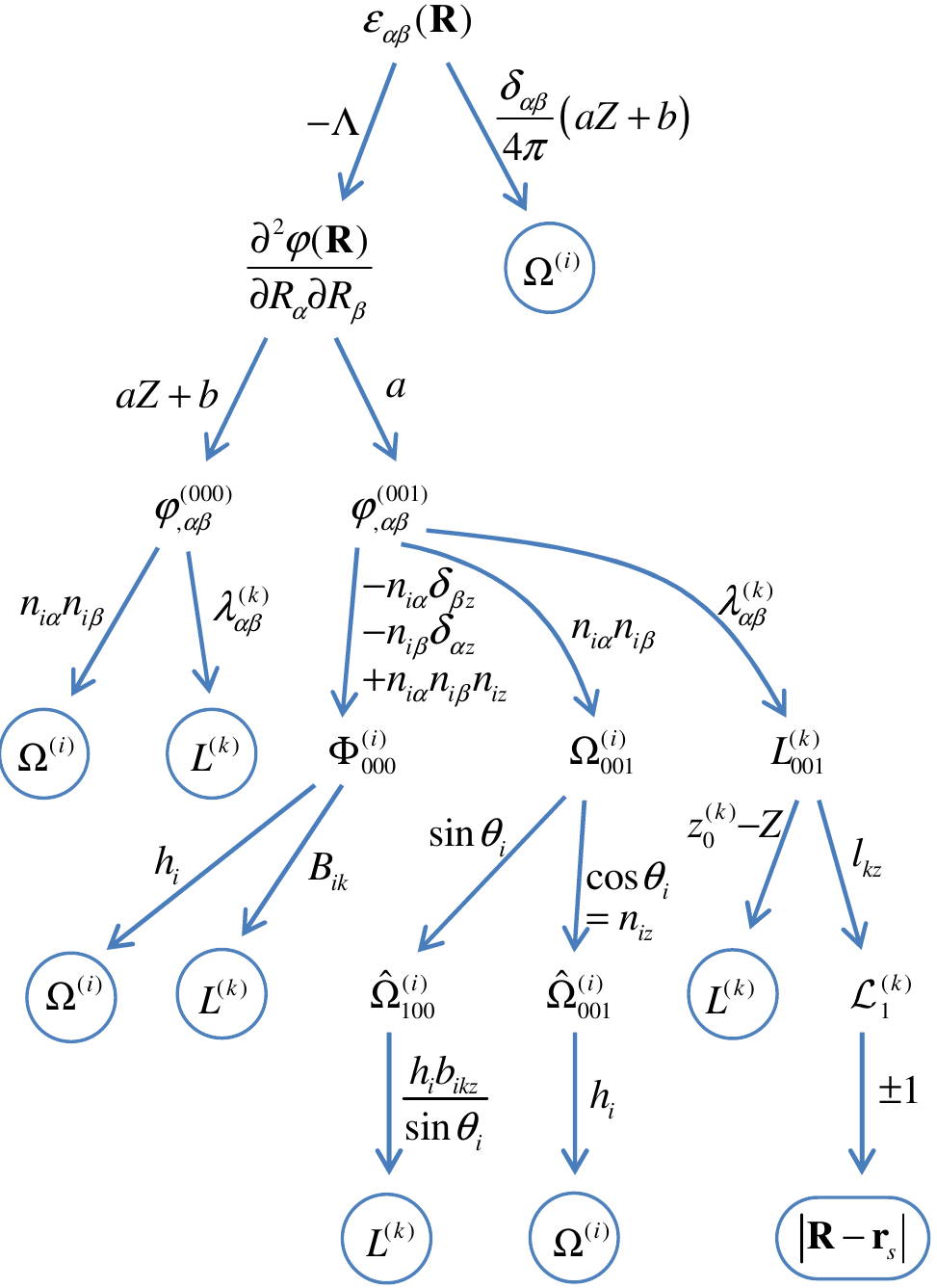}
  \caption{Schematic representation of the algorithm of calculating 
  the strain tensor $\ve_{\alpha\beta}$  
  in the case of a linearly varied eigenstrain: $f=az+b$.}
  \label{fig:graph-linear-f}
\end{figure}

Now let us examine the case of linear dependence of the eigenstrain $\ve_0$ 
on the coordinates inside the inclusion. 
For simplicity, we suppose in this Subsection that the eigenstrain depends 
on coordinate $z$ only:
\begin{equation}
  f(x,y,z) = az + b ,
\end{equation}
where $a$ (the gradient of the eigenstrain) and $b$ are constants. 
For such a function $f(\r)$, the algorithm of Section~\ref{sec:main} 
is shown in a graphical form in Fig.~\ref{fig:graph-linear-f}. The meaning of arrows 
and expressions in the nodes of the graph is explained above, see the caption 
to Fig.~\ref{fig:graph-constant-f}. 
In the part of the tree below $\Omega^{(i)}_{001}$, we use the results 
of Subsection~\ref{sec:reducing-example}. 
For the arrow connecting $\mathcal{L}^{(k)}_1$ to $|\R-\r_s|$, the factor is 
equal to $+1$ if the vector $\vec{l}_k$ is directed from the opposite vertex 
of $k$th edge to $s$th vertex, and $-1$ otherwise.

The expressions for the polynomials $\mathcal{A}^{(i)}_{\ve\alpha\beta}$, 
$\mathcal{B}^{(k)}_{\ve\alpha\beta}$, $\mathcal{C}^{(s)}_{\ve\alpha\beta}$ 
can be derived from the graph in Fig.~\ref{fig:graph-linear-f} by the method 
described in Subsection~\ref{sec:examples-homogeneous}. The only new feature 
is the existence of a node $|\R-\r_s|$ that provides a contribution to 
the polynomial $\mathcal{C}^{(s)}_{\ve\alpha\beta}$, which is equal to 
the product of the 
quantities near the arrows on the way from $\ve_{\alpha\beta}$ to $|\R-\r_s|$.
As a result, one can get the following answers for 
$\mathcal{A}^{(i)}_{\ve\alpha\beta}$, 
$\mathcal{B}^{(k)}_{\ve\alpha\beta}$ and $\mathcal{C}^{(s)}_{\ve\alpha\beta}$:
\begin{subequations}
\begin{multline} \label{linear-z-A}
  \mathcal{A}^{(i)}_{\ve\alpha\beta}(\R) 
  = - \Lambda (aZ+b) n_{i\alpha} n_{i\beta} \\
  - \Lambda a \left( - n_{i\alpha}\delta_{\beta z} 
  - n_{i\beta}\delta_{\alpha z} + 2 n_{i\alpha} n_{i\beta} n_{iz} \right) h_i 
  + (aZ+b) \, \frac{\delta_{\alpha\beta}}{4\pi} \, ,
\end{multline}
\begin{multline} \label{linear-z-B}
  \mathcal{B}^{(k)}_{\ve\alpha\beta}(\R) 
  = - \Lambda (aZ+b) \lambda^{(k)}_{\alpha\beta} \\
  - \Lambda a \sum\limits_i \left( - n_{i\alpha}\delta_{\beta z} 
  - n_{i\beta}\delta_{\alpha z} + n_{i\alpha} n_{i\beta} n_{iz} \right) B_{ik} \\
  - \Lambda a \sum\limits_i n_{i\alpha} n_{i\beta} b_{ikz} h_i 
  - \Lambda a \lambda^{(k)}_{\alpha\beta} \big( z^{(k)}_0-Z \big) ,
\end{multline}
where index $i$ runs over the two faces adjacent to $k$th edge,
\begin{equation} \label{linear-z-C}
  \mathcal{C}^{(s)}_{\ve\alpha\beta}(\R) 
  = - \Lambda a \sum\limits_k \lambda^{(k)}_{\alpha\beta} (\pm l_{kz}) ,
\end{equation}
\end{subequations}
where index $k$ runs over the edges that enter $s$th vertex. 
The sign in the factor $(\pm l_{kz})$ 
is chosen to be plus if the vector $\vec{l}_k$ 
is directed from the opposite vertex of $k$th edge 
to $s$th vertex, and minus otherwise. 
Constants $\Lambda$, $\lambda^{(k)}_{\alpha\beta}$, 
and linear functions $h_i(\R)$, $B_{ik}(\R)$, $z^{(k)}_0(\R)$ 
are defined by Eqs.~\eq{Lambda-via-nu}, 
\eq{lambda-k-alpha-beta}, \eq{h-i}, \eq{B-ik}, \eq{r0}, correspondingly.

In the next Subsection, these equations will be generalized to the case of an arbitrarily 
directed gradient of the lattice misfit.

\subsection{Constant gradient of the lattice misfit: general case}
\label{sec:examples-linear}

Consider the general case of a constant gradient of the lattice misfit 
inside the polyhedral inclusion:
\begin{equation}
  f(\r) = \vec{a} \cdot \r + b ,
\end{equation}
where a vector $\vec{a}$ and a scalar $b$ are constants. It is easy to generalize 
formulas~\eq{linear-z-A}--\eq{linear-z-C} to this case. For this purpose, it is enough to 
represent them in a coordinate-independent form, i.~e., to replace the product $aZ$ 
by $\vec{a} \cdot \R$, the product $an_{iz}$ by $\vec{a} \cdot \vec{n}_i$, etc. 
This gives rise to the following result:
\begin{subequations} \label{linear}
\begin{multline} \label{linear-A}
  \mathcal{A}^{(i)}_{\ve\alpha\beta}(\R) 
  = - \Lambda n_{i\alpha} n_{i\beta} \left[ \vec{a}\cdot\R + b + 2(\vec{a}\cdot\vec{n}_i) h_i \right] \\
  + \Lambda ( a_\alpha n_{i\beta} + a_\beta n_{i\alpha} ) h_i 
  + (\vec{a}\cdot\R + b) \, \frac{\delta_{\alpha\beta}}{4\pi} \, ,
\end{multline}
\begin{multline} \label{linear-B}
  \mathcal{B}^{(k)}_{\ve\alpha\beta}(\R) 
  = - \Lambda \lambda^{(k)}_{\alpha\beta} \left( \vec{a}\cdot\r^{(k)}_0 + b \right) \\
  + \Lambda \sum\limits_i ( a_\alpha n_{i\beta} + a_\beta n_{i\alpha} ) B_{ik} \\
  - \Lambda \sum\limits_i n_{i\alpha} n_{i\beta} \left[ (\vec{a}\cdot\vec{n}_i) B_{ik} + (\vec{a}\cdot\vec{b}_{ik}) h_i \right] ,
\end{multline}
\begin{equation} \label{linear-C}
  \mathcal{C}^{(s)}_{\ve\alpha\beta}(\R) 
  = - \Lambda \sum\limits_k \lambda^{(k)}_{\alpha\beta} ( \pm \vec{a}\cdot\vec{l}_k ) .
\end{equation}
\end{subequations}
In order to get the strain distribution $\ve_{\alpha\beta}(\R)$, one should insert 
these expressions into the universal equation~\eq{main-result}.

By the same method, one can get the analytical formulas for the potential $\phi$ 
(as well as its first and second derivatives) of a non-uniformly charged polyhedral 
body with a charge density $\vec{a} \cdot \r + b$. The polynomials 
$\mathcal{A}^{(i)}_\phi$, $\mathcal{B}^{(k)}_\phi$ and $\mathcal{C}^{(s)}_\phi$ 
related to the potential are expressed as follows:
\begin{subequations} \label{linear-phi}
\begin{equation} \label{linear-A-phi}
  \mathcal{A}^{(i)}_\phi(\R) 
  = \left(  \frac{\vec{a}\cdot\R + b}{2} + \frac{ (\vec{a}\cdot\vec{n}_i) h_i }{3}  \right)  h_i^2 \, ,
\end{equation}
\begin{multline} \label{linear-B-phi}
  \mathcal{B}^{(k)}_\phi(\R) = 
    \frac16 \sum\limits_i \left[ \vec{a} \cdot \big(  2\R + \vec{n}_i h_i + \r^{(k)}_0  \big) + 3b \right] h_i B_{ik} \\
  + \frac16 \sum\limits_i (\vec{a}\cdot\vec{b}_{ik}) h_i^3 ,
\end{multline}
\begin{equation} \label{linear-C-phi}
  \mathcal{C}^{(s)}_\phi(\R) 
  = \frac16 \sum\limits_k \left[  ( \pm \vec{a}\cdot\vec{l}_k )  \sum\limits_i h_i B_{ik} \right] .
\end{equation}
\end{subequations}
Eqs.~\eq{linear-A-phi}--\eq{linear-C-phi} are to be inserted 
into Eq.~\eq{main-result-phi} in order to get the potential $\phi(\R)$.

\begin{figure}
  \includegraphics[width=\columnwidth]{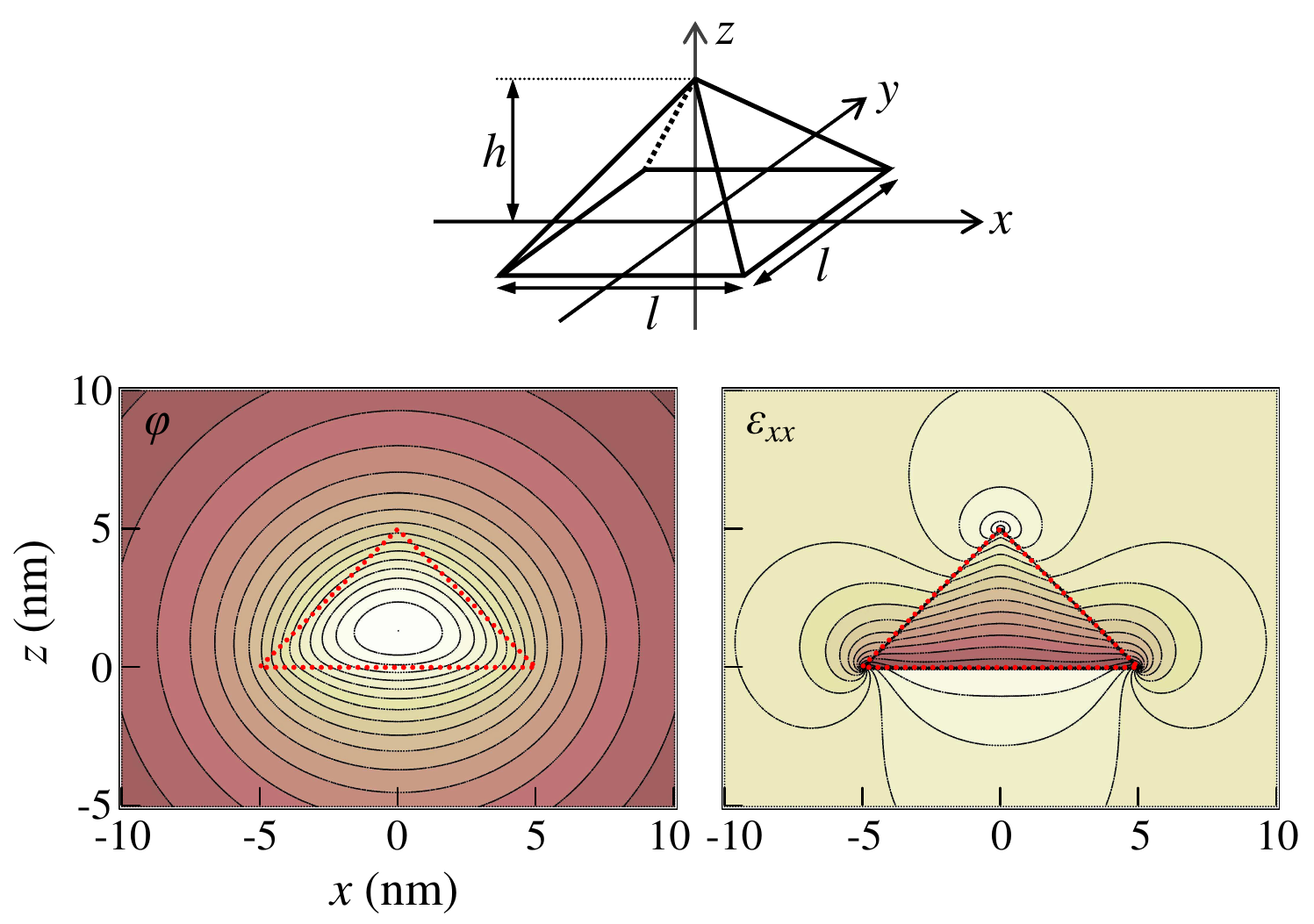}
  \caption{Isoline plots (in the plane $y=0$) of the potential $\phi(\r)$ 
  and the strain component $\ve_{xx}(\r)$ 
  produced by a pyramidal inclusion with height $h=5$~nm 
  and the length of the square base $l=10$~nm. The eigenstrain $\ve_0(\r)$ inside 
  the inclusion decreases linearly from the base to the apex: $\ve_0 \propto (1-z/2h)$. 
  Poisson ratio $\nu = 0.25$. 
  The potential is calculated via Eqs.~\eq{main-result-phi}, \eq{linear-phi}, 
  and the strain---via Eqs.~\eq{main-result}, \eq{linear}. 
  The isolines are equidistant in $\phi$ and $\ve_{xx}$, brighter areas correspond to larger values. 
  The inclusion cross section is shown by the red dotted contour.}
  \label{fig:linear-results}
\end{figure}
Analytical expressions for the gravitational potential 
of a polyhedral body having a linearly varying density distribution 
were reported by many authors.\cite{Pohanka1998,Holstein2003, Hamayun2009, DUrso2014, Conway2015} 
However these expressions were not represented in terms of solid angles. 
For this reason, 
it seems to be extremely difficult to compare them directly with our answer 
[a combination of Eqs.~\eq{main-result-phi} and~\eq{linear-phi}] 
at the level of formulas. 
To ensure that the results of this Subsection are correct, we performed 
numerical tests, considering a pyramid with a square base as an example 
of an inclusion (see Fig.~\ref{fig:linear-results}). The following statements 
were checked numerically:
\begin{itemize}
\item $\phi(x,y,z)$ is a continuous function, and its first derivatives are continuous;
\item $\phi(x,y,z)$ obeys Poisson's equation $\Delta\phi = -4\pi\ve_0$, where 
$\ve_0(\r) = \vec{a} \cdot \r + b$ inside the pyramid, and $\ve_0=0$ 
outside the pyramid; 
\item the functions $\phi(x,y,z)$ and $\ve_{\alpha\beta}(x,y,z)$ are connected 
to each other by Eq.~\eq{epsilon-via-potential}.
\end{itemize}
The potential $\phi$ and the strain $\ve_{\alpha\beta}$ were calculated by 
Eqs.~\eq{main-result-phi},~\eq{linear-phi} and Eqs.~\eq{main-result},~\eq{linear}, 
correspondingly. The pyramid height $h$, the length $l$ of the edge of the pyramid base, 
the gradient $\vec{a}$ of the lattice misfit, and the free term $b$ were varied 
independently. An example of the calculated potential $\phi$ and $xx$-component 
of the strain is shown in Fig.~\ref{fig:linear-results} for the case of $h=5$~nm, $l=10$~nm, $a_x=a_y=0$, 
and $a_z=-b/2h$. (Such a choice of $\vec{a}$ means that the lattice misfit 
linearly decreases from the base to the apex of the pyramid, so that at the apex 
it is twice smaller than at the base.) 
For this case, as well as for many other combinations of parameters 
$h$, $l$, $\vec{a}$ and $b$, the numerical test has been passed successfully.

In order to provide more evidences of validity of analytical results, we have compared 
the distribution of $\ve_{xx}$ calculated analytically and shown in Fig.~\ref{fig:linear-results} 
with analogous distributions calculated numerically using the finite element method. 
This comparison is presented in Appendix~\ref{sec:comsol}. 
One can see that for large enough size of the ``box'', in which the inclusion is incorporated, 
the numerical results perfectly agree with the analytical ones.

\subsection{Sinusoidal profile of the lattice misfit}
\label{sec:examples-sinusoidal}

As an example of a more complicated distribution of the lattice misfit $\ve_0(x,y,z)$, 
let us consider the sinusoidal profile inside the inclusion:
\begin{equation} \label{f-sin}
  \ve_0(\r) = 
  \begin{cases}
    A \sin(kx) & \text{if } \r \text{ is inside the inclusion}, \\
	0 & \text{if } \r \text{ is outside the inclusion},
  \end{cases}
\end{equation}
so that $f(\r) = A \sin(kx)$, where $A$ is a constant. 
In order to apply the results of the present paper, one has to approximate 
the function $f(\r)$ by a polynomial. As such an approximation, we choose 
first five terms of the Fourier series,
\begin{equation} \label{f-sin-approx}
  f(\r) \approx A \left[ kx - \frac{(kx)^3}{3!} + \frac{(kx)^5}{5!} - \frac{(kx)^7}{7!} + \frac{(kx)^9}{9!} \right] ,
\end{equation}
that provides accuracy not worse than one percent of $A$ within one period, i.~e., 
for $|kx|\leq\pi$.

\begin{figure}
  \includegraphics[width=\columnwidth]{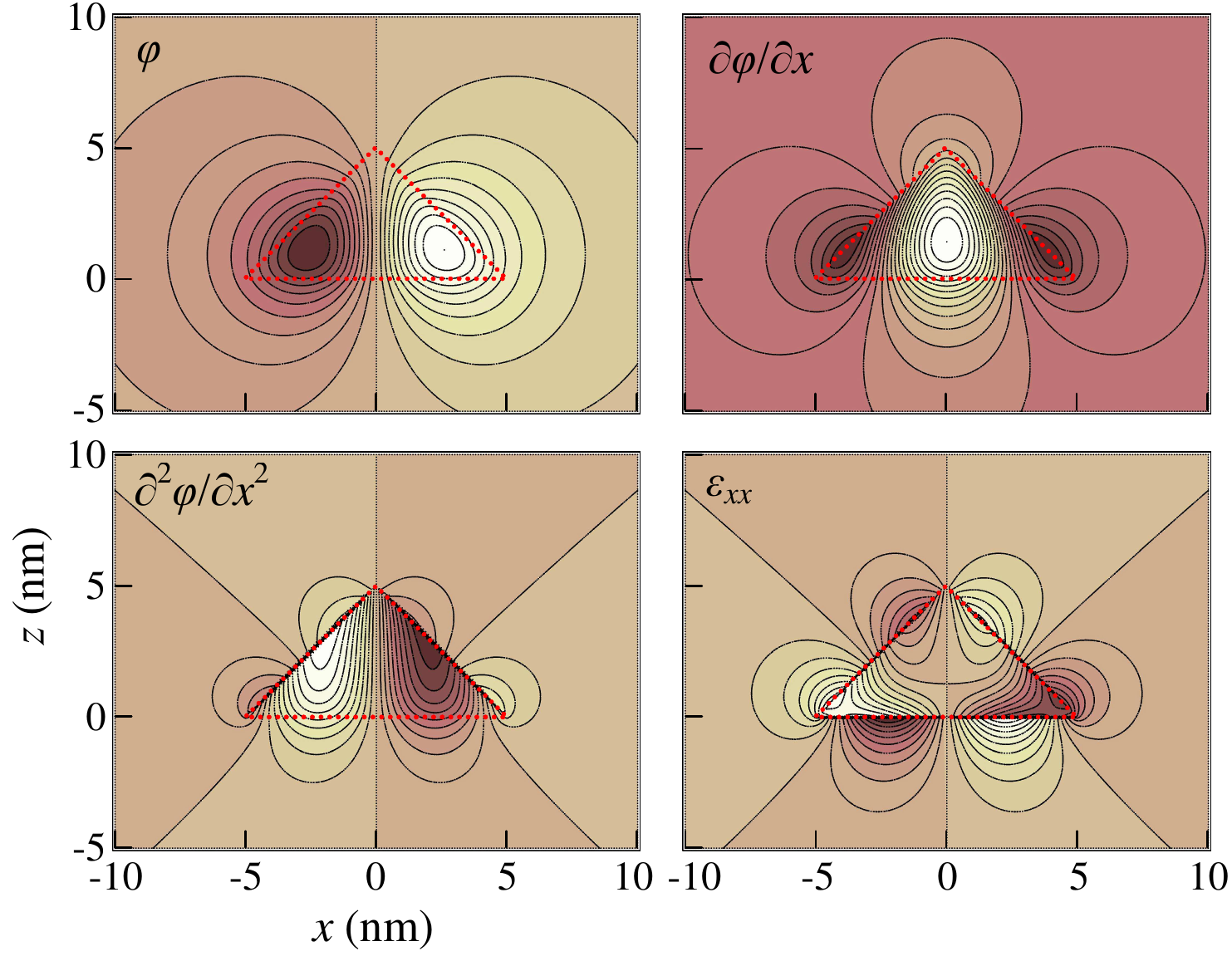}
  \caption{Isoline plots (in the plane $y=0$) of the potential~$\phi$, its derivatives 
  $\d\phi/\d x$, $\d^2\phi/\d x^2$, 
  and the strain component $\ve_{xx}$   produced by a pyramidal inclusion with height $h=5$~nm 
  and the length of the square base $l=10$~nm. 
  (See Fig.~\ref{fig:linear-results} for a sketch of the inclusion.) 
  The eigenstrain $\ve_0(\r)$ inside 
  the inclusion is approximately proportional to $\sin(kx)$ with $k=2\pi/l$, see Eq.~\eq{f-sin-approx}. 
  Poisson ratio $\nu = 0.25$. 
  The isolines are equidistant, brighter areas correspond to larger values. 
  The inclusion cross section is shown by the red dotted contour.}
  \label{fig:sinus-results}
\end{figure}

In Fig.~\ref{fig:sinus-results}, an example of the potential $\phi(x,y,z)$, its 
derivatives $\d\phi/\d x$ and $\d^2\phi/\d x^2$, and the $xx$-component of the 
strain tensor is shown for the same pyramidal inclusion 
as in Fig.~\ref{fig:linear-results}. The height $h$ and the lateral size $l$ 
of the pyramid are 5~nm and 10~nm, correspondingly. The parameter $k$ is chosen to 
be $k=2\pi/l$, so that the pyramid size $l$ corresponds to the period of the sinusoid. 

The values of the strain tensor, the potential, and its first and second derivatives 
were calculated by Eqs.~\eq{main-result}--\eq{main-result-phi-2deriv}. 
The coefficients of polynomials $\mathcal{A}$, $\mathcal{B}$, $\mathcal{C}$, 
which contribute to these equations, were calculated numerically, directly 
following the algorithm of Subsection~\ref{sec:main-algorithm}. 
Then, the same numerical test as for a linearly varying eigenstrain 
(see the previous Subsection) was performed and successfully passed. In addition, 
it was checked that the results for the potential and its first and second 
derivatives are consistent with each other.

Such a numerical test with a degree $N$ of the polynomial $f(\r)$, larger than 1, 
is important for the following reason. Equations~\eq{Omega-reduce-n}, 
\eq{Omega-reduce-m} and \eq{L-n-answer} are used at Steps~8, 9, 11 of the algorithm 
only when index $m$, $n$ or $t$ becomes larger than 1, that occurs 
only if $N\geq2$.  Hence, to check the algorithm, it is necessary 
to consider at least quadratic ($N=2$) coordinate dependence of the eigenstrain. 
In the example of this Subsection, $N$ is equal to 9, that provides a good test 
of the algorithm as a whole.

\section{Concluding remarks}
\label{sec:conclusions}

Two problems, closely related to each other, are considered in this paper. 
The first one is the search for an analytical solution (in a closed form) for the spatial distribution 
of the elastic strain induced by a lattice-mismatched inclusion inserted in an infinite matrix, 
in the case of a polyhedral shape of the inclusion. 
The second problem is finding an analytical solution for an electrostatic of gravitational 
potential created by a polyhedral body, and for the derivatives of the potential. 
It is well known that the strain tensor is expressed via the second derivatives of the potential, 
see Eq.~\eq{epsilon-via-potential}, assuming that the inclusion and the matrix are 
elastically isotropic and have the same elastic constants. Under these assumptions, 
we obtained the universal analytical expression describing the strain distribution, 
Eq.~\eq{main-result}, for a polynomial coordinate dependence of the lattice misfit within the inclusion. 
Similar analytical formulas, Eqs.~\eq{main-result-phi}--\eq{main-result-phi-2deriv}, were derived 
for the potential and its first and second derivatives in the case of a polynomial distribution of 
the charge/mass density over a polyhedral-shaped body. 

The results of the present work allow one to understand the structure of the known analytical solutions for 
different particular cases, e. g. strain distributions due to homogeneous polyhedral inclusions, 
potentials of polyhedral bodies with a constant density or a constant gradient of the density. 
Typically, the solutions found in the literature are combinations of logarithmic terms and terms containing 
inverse trigonometric functions, mostly arctangents. Our expressions~\eq{main-result-phi}--\eq{main-result-phi-2deriv} 
clarify that logarithmic terms arise from potentials of uniformly charged edges $L^{(k)}$, and arctangent 
terms---from solid angles $\Omega^{(i)}$.

Finally, it is worth highlighting several ways of generalizing the results of the present study. The strain distribution 
for an inclusion in a half-space can be found by the ``mirror image'' method.\cite{Davies2003} 
The effects of anisotropy can be taken into account either by series expansion of the elastic Green's function,\cite{Faux2000, Faux2005} 
or by the method of stretching the coordinates.\cite{Kuvshinov2008, Nenashev2013en} 
In the case of different elastic constants of the matrix and the inclusion (an inhomogeneous inclusion), 
one can apply the perturbation method based on the concept of effective inclusion.\cite{Chu2011, Ma2014}

\acknowledgments This work is funded by RFBR (grants \#16-29-14031 and \#16-02-00397, development of analytical expressions) 
and by State Programme (Grant No. 0306-2016-0015, creating the computer program for calculation of the strain).


\appendix

\section{Integrals containing the derivatives of characteristic function $\chi(\r)$}
\label{app:chi}

This Appendix is devoted to reducing the integrals $\phi^{(mnp)}_{,\alpha}$ 
and $\phi^{(mnp)}_{,\alpha\beta}$ 
defined by Eqs.~\eq{phi-mnp-alpha} and~\eq{phi-mnp-alpha-beta} to the primitive introduced in Section~\ref{sec:primitives}. 

Let us define a function $f(\r)$ as
\begin{equation}
  f(x,y,z) = (x-X)^m (y-Y)^n (z-Z)^p \, ,
\end{equation}
then
\begin{equation} \label{app-phi-mnp-alpha}
  \phi^{(mnp)}_{,\alpha} 
  = \iiint \frac{\d[f(\r)\,\chi(\r)]}{\d r_\alpha} \; \frac{d^3r}{|\r-\R|}
\end{equation}
and
\begin{equation} \label{app-phi-mnp-alpha-beta}
  \phi^{(mnp)}_{,\alpha\beta} 
  = \iiint \frac{\d^2[f(\r)\,\chi(\r)]}{\d r_\alpha\d r_\beta} \; \frac{d^3r}{|\r-\R|} \, .
\end{equation}

First we will focus on the integral $\phi^{(mnp)}_{,\alpha}$. Let us open the brackets in the expression $\d(f\chi) / \d r_\alpha$:
\begin{equation}
  \frac{\d(f\chi)}{\d r_\alpha} 
  = \frac{\d f}{\d r_\alpha} \, \chi 
  + f \, \frac{\d \chi}{\d r_\alpha} \, .
\end{equation}
Substituting this expansion into Eq.~\eq{app-phi-mnp-alpha}, one can see that $\phi^{(mnp)}_{,\alpha}$ is equal to the sum of two integrals, 
the first of which being the quantity $\phi_{(mnp),\alpha}$ defined by Eq.~\eq{phi-low-mnp-alpha}. Therefore 
\begin{equation} \label{app-phi-up-phi-down}
  \phi^{(mnp)}_{,\alpha} 
  = \phi_{(mnp),\alpha}
  + \iiint f \, \frac{\d \chi}{\d r_\alpha} \; \frac{d^3r}{|\r-\R|} \, .
\end{equation}
In the remaining integral, the derivative $\d \chi / \d r_\alpha$ differs from zero only on the surface of the polyhedron; 
thus, it is in fact a surface integral. To be more precise, let us assume that the polyhedron is \emph{convex} 
(generalization to non-convex polyhedra does not lead to any difficulties), and define the characteristic function $\chi(\r)$ as follows:
\begin{equation} \label{app-chi}
  \chi(\r) = \prod\limits_{\substack{i \\ \text{(faces)}}} \theta\big[ \vec{n}_i \cdot (\r_i-\r) \big] ,
\end{equation}
where $\theta$ denotes the Heaviside step function, $\vec{n}_i$ is the outward normal to $i$th face, and $\r_i$ is a position vector of some 
(arbitrarily chosen) point on $i$th face. It is easy to check that Eqs.~\eq{chi} and~\eq{app-chi} define the same function $\chi(\r)$. 
Differentiation of each factor in Eq.~\eq{app-chi} is straightforward:
\begin{equation} \label{app-d-theta-dr}
  \frac{\d}{\d r_\alpha} \theta\big[ \vec{n}_i \cdot (\r_i-\r) \big] = - n_{i\alpha} \delta\big[ \vec{n}_i \cdot (\r_i-\r) \big] ,
\end{equation}
where $\delta$ denotes the Dirac delta function. Using Eq.~\eq{app-d-theta-dr}, one can represent the derivative $\d \chi / \d r_\alpha$ 
as follows:
\begin{equation} \label{app-d-chi-dr}
  \frac{\d \chi}{\d r_\alpha} = - \sum\limits_i n_{i\alpha} \psi_i(\r) ,
\end{equation}
where functions $\psi_i$ are defined as
\begin{equation} \label{app-psi-def}
  \psi_i(\r) = \delta\big[ \vec{n}_i \cdot (\r_i-\r) \big] \prod\limits_{j \neq i} \theta\big[ \vec{n}_j \cdot (\r_j-\r) \big] .
\end{equation}
One can easily see that the function $\psi_i$ vanishes everywhere except $i$th face 
of the polyhedron's surface, where it has a $\delta$-like singularity. Hence, 
any volume integral containing the function $\psi_i$ is in fact a surface integral 
over $i$th face. In particular,
\begin{equation} \label{app-int-psi}
  \iiint f \psi_i \, \frac{d^3r}{|\r-\R|} 
  = \iint\limits_{\text{face} \, i} f \, \frac{dS}{|\r-\R|} 
  = \Phi^{(i)}_{mnp} \, .
\end{equation}
Using Eqs.~\eq{app-d-chi-dr} and~\eq{app-int-psi}, one can reduce the integral in 
the right-hand side of Eq.~\eq{app-phi-up-phi-down} 
to surface integrals $\Phi^{(i)}_{mnp}$:
\begin{equation}
  \iiint f \, \frac{\d \chi}{\d r_\alpha} \; \frac{d^3r}{|\r-\R|} 
  = - \sum\limits_i n_{i\alpha} \Phi^{(i)}_{mnp} .
\end{equation}
Substitution of this representation to Eq.~\eq{app-phi-up-phi-down} gives rise to 
Eq.~\eq{phi-alpha-via-phi-Phi}.

Hence, we have proved Eq.~\eq{phi-alpha-via-phi-Phi}, i.~e., have found out 
a representation of the integral $\phi^{(mnp)}_{,\alpha}$ in terms of primitives.

Now let us consider the integral $\phi^{(mnp)}_{,\alpha\beta}$. 
The second derivative in its definition, Eq.~\eq{app-phi-mnp-alpha-beta}, can be 
expanded as follows:
\begin{equation} \label{app-d2-f-chi-dr-dr}
  \frac{\d^2(f\chi)}{\d r_\alpha \d r_\beta} 
  = \frac{\d^2 f}{\d r_\alpha \d r_\beta} \, \chi 
  + \frac{\d f}{\d r_\beta} \, \frac{\d \chi}{\d r_\alpha}
  + \frac{\d f}{\d r_\alpha} \, \frac{\d \chi}{\d r_\beta}
  + f \, \frac{\d^2 \chi}{\d r_\alpha \d r_\beta} .
\end{equation}
After substituting this expansion to Eq.~\eq{app-phi-mnp-alpha-beta}, one gets 
four integrals corresponding to four terms of the right-hand side 
of Eq.~\eq{app-d2-f-chi-dr-dr}. The first integral is equal to the quantity 
$\phi_{(mnp),\alpha\beta}$ introduced in Eq.~\eq{phi-low-mnp-alpha-beta}:
\begin{equation}
  \iiint \frac{\d^2 f}{\d r_\alpha \d r_\beta} \, \chi \, \frac{d^3r}{|\r-\R|}
  = \phi_{(mnp),\alpha\beta} \, .
\end{equation}
The second integral, which contains the derivative $\d \chi/\d r_\alpha$, 
can be treated as above, via Eqs.~\eq{app-d-chi-dr} and~\eq{app-int-psi}:
\begin{multline}
  \iiint \frac{\d f}{\d r_\beta} \; \frac{\d \chi}{\d r_\alpha} \; \frac{d^3r}{|\r-\R|} 
  = - \sum\limits_i n_{i\alpha} \iiint \frac{\d f}{\d r_\beta} \, \psi_i \, \frac{d^3r}{|\r-\R|} \\
  = - \sum\limits_i n_{i\alpha} \iint\limits_{\text{face} \, i} \frac{\d f}{\d r_\beta} \; \frac{dS}{|\r-\R|} 
  = - \sum\limits_i n_{i\alpha} \Phi^{(i)}_{(mnp),\beta} \, ,
\end{multline}
where $\Phi^{(i)}_{(mnp),\beta}$ is the quantity defined by Eq.~\eq{Phi-mnp-alpha}. 
The third integral is the same as the second one, up to the index permutation 
$\alpha\leftrightarrow\beta$. Therefore the expression for $\phi^{(mnp)}_{,\alpha\beta}$ 
takes the following form:
\begin{multline} \label{app-phi-mnp-alpha-beta-1}
  \phi^{(mnp)}_{,\alpha\beta}  
  = \phi_{(mnp),\alpha\beta} 
  - \sum\limits_i n_{i\alpha} \Phi^{(i)}_{(mnp),\beta} 
  - \sum\limits_i n_{i\beta} \Phi^{(i)}_{(mnp),\alpha} \\
  + \iiint f \, \frac{\d^2 \chi}{\d r_\alpha \d r_\beta} \; \frac{d^3r}{|\r-\R|} \, .
\end{multline}
In order to go further, one needs a method of dealing with integrals containing 
the second derivative $\d^2 \chi / \d r_\alpha \d r_\beta$. Using Eq.~\eq{app-d-chi-dr}, 
this second derivative can be represented~as
\begin{equation} \label{app-d2chi-dr2}
  \frac{\d^2 \chi}{\d r_\alpha \d r_\beta} 
  = - \sum\limits_i n_{i\alpha} \, \frac{\d \psi_i}{\d r_\beta} .
\end{equation}
Then, it is more convenient to define the function $\psi_i(\r)$ in an alternative way:
\begin{equation} \label{app-psi-def2}
  \psi_i(\r) = \delta\big[ \vec{n}_i \cdot (\r_i-\r) \big] 
  \prod\limits_{\substack{k \\ \text{(edges)}}} \theta\big[ \vec{b}_{ik} \cdot (\r_k-\r) \big],
\end{equation}
where index $k$ runs over the edges that surround $i$th face, and unit vectors 
$\vec{b}_{ik}$ are directed out of $i$th face perpendicularly to their corresponding 
edges, as shown in Fig.~\ref{fig:n-b}. One can easily figure out that both 
equations~\eq{app-psi-def} and~\eq{app-psi-def2} define the same function, which 
has delta-function-like behaviour on $i$th face and is equal to zero outside this face. 
Consequently, the derivative $\d \psi_i/\d r_\beta$ takes the form
\begin{equation} \label{app-d-psi-dr}
  \frac{\d \psi_i}{\d r_\beta} 
  = - n_{i\beta} \omega_i(\r) - \sum\limits_k b_{ik\beta} \eta_k(\r) ,
\end{equation}
where
\begin{equation}
  \omega_i(\r) = \delta'\big[ \vec{n}_i \cdot (\r_i-\r) \big] 
  \prod\limits_k \theta\big[ \vec{b}_{ik} \cdot (\r_k-\r) \big]
\end{equation}
and
\begin{equation} \label{app-eta}
  \eta_k(\r) = \delta\big[ \vec{n}_i \cdot (\r_i-\r) \big] 
  \delta\big[ \vec{b}_{ik} \cdot (\r_k-\r) \big] 
  \prod\limits_{k' \neq k} \theta\big[ \vec{b}_{ik'} \cdot (\r_{k'}-\r) \big] .
\end{equation}
(Notably, the function $\eta_k(\r)$ is one and the same for both faces adjacent to 
$k$th edge.) The term containing $\omega_i$ arises in Eq.~\eq{app-d-psi-dr} 
from differentiating the delta-functional factor in Eq.~\eq{app-psi-def2}, 
and the terms with $\eta_k$---from differentiating Heaviside $\theta$-functions 
in a manner similar to Eq.~\eq{app-d-theta-dr}. 
With Eqs.~\eq{app-d2chi-dr2} and~\eq{app-d-psi-dr} one can represent second derivatives 
of $\chi$ as
\begin{equation}
  \frac{\d^2 \chi}{\d r_\alpha \d r_\beta} 
  = \sum\limits_i n_{i\alpha} \, n_{i\beta} \, \omega_i(\r) 
  + \sum\limits_i \sum\limits_k n_{i\alpha} \, b_{ik\beta} \, \eta_k(\r) ,
\end{equation}
where index $i$ runs over all faces, and index $k$---over edges adjacent to face $i$. 
Changing the order of summations in the last sum, and introducing the tensor 
$\lambda^{(k)}_{\alpha\beta}$ in accordance with Eq.~\eq{lambda-k-alpha-beta}, 
one can rewrite this representation in a more convenient form:
\begin{equation} \label{app-d2chi-dr2-via-omega-eta}
  \frac{\d^2 \chi}{\d r_\alpha \d r_\beta} 
  = \sum\limits_i n_{i\alpha} \, n_{i\beta} \, \omega_i(\r) 
  + \sum\limits_k \lambda^{(k)}_{\alpha\beta} \, \eta_k(\r) ,
\end{equation}
where index $k$ runs over all edges of the polyhedron. Hence, the integral in the 
right-hand side of Eq.~\eq{app-phi-mnp-alpha-beta-1} is a linear combination of 
similar integrals containing $\omega_i$ or $\eta_k$ 
instead of $\d^2 \chi / \d r_\alpha \d r_\beta$.

Let us consider the integral with $\omega_i$. For convenience, we temporarily assume 
that $i$th face is perpendicular to the coordinate axis $z$. Then, the function 
$\omega_i(\r)$ can be expressed as
\begin{equation}
  \omega_i(x,y,z) = \delta'(z_i-z) \, \tilde\chi_i(x,y) ,
\end{equation}
where $z_i$ is the value of $z$ at $i$th face, and the function $\tilde\chi_i(x,y)$ 
defines the polygon of $i$th face: $\tilde\chi_i(x,y)=1$ if the point $(x,y,z_i)$ 
lies inside the face, otherwise $\tilde\chi_i(x,y)=0$. Therefore one can 
represent the integral with $\omega_i$ as follows:
\begin{multline} \label{app-int-with-omega-1}
  \iiint f \omega_i \, \frac{d^3r}{|\r-\R|} \\
  = \iint \tilde\chi_i(x,y) \, dx\,dy \int \delta'(z_i-z) \, \frac{f}{|\r-\R|} \, dz .
\end{multline}
The inner integral can be taken by parts,
\begin{multline} \label{app-int-with-omega-2}
  \int \delta'(z_i-z) \, \frac{f}{|\r-\R|} \, dz 
  = \int \delta(z_i-z) \, \frac{\d}{\d z} \left(\frac{f}{|\r-\R|}\right) \, dz \\
  = \int \delta(z_i-z) \, \left( \frac{\d f}{\d z} - f\frac{z-Z}{|\r-\R|^2} \right) 
  \frac{dz}{|\r-\R|}  \, ,
\end{multline}
and substitution of Eq.~\eq{app-int-with-omega-2} into Eq.~\eq{app-int-with-omega-1} 
leads to the following result:
\begin{multline} \label{app-int-with-omega-3}
  \iiint f \omega_i \, \frac{d^3r}{|\r-\R|} 
  = \iint\limits_{\text{face} \, i} 
  \left( \frac{\d f}{\d z} - f\frac{z-Z}{|\r-\R|^2} \right)  \frac{dS}{|\r-\R|} \\
  = \Phi^{(i)}_{(mnp),z} + \Omega^{(i)}_{mnp} \, ,
\end{multline}
where the quantity $\Phi^{(i)}_{(mnp),z}$ is defined in Eq.~\eq{Phi-mnp-alpha}. 
Though Eq.~\eq{app-int-with-omega-3} was derived in a special coordinate frame, 
one can easily make it independent of the choice of the frame. For that, it is enough 
to rewrite the term $\Phi^{(i)}_{(mnp),z}$ as $n_{i\gamma} \Phi^{(i)}_{(mnp),\gamma}$, 
i. e., in the invariant manner:
\begin{equation} \label{app-int-with-omega}
  \iiint f \omega_i \, \frac{d^3r}{|\r-\R|} 
  = n_{i\gamma} \Phi^{(i)}_{(mnp),\gamma} + \Omega^{(i)}_{mnp} \, .
\end{equation}

In order to calculate a similar integral with $\eta_k$, we choose (temporarily) 
a coordinate frame with axis $z$ along the vector $\vec{n}_i$, and axis $x$ 
along the vector $\vec{b}_{ik}$, so that axis $y$ is directed along $k$th edge. 
In this frame, the expression~\eq{app-eta} can be rewritten as
\begin{equation}
  \eta_k(x,y,z) = \delta(z_k-z) \, \delta(x_k-x) \, \tilde\chi_k(y) ,
\end{equation}
where $x_k$ and $z_k$ are coordinates of $k$th edge, and the function 
$\tilde\chi_k(y)$ is equal to 1 if the point $(x_k,y,z_k)$ lies on the edge, 
otherwise $\tilde\chi_k(y)=0$. Then it is obvious that
\begin{equation} \label{app-int-with-eta}
  \iiint f \eta_k \, \frac{d^3r}{|\r-\R|} = L^{(k)}_{mnp}
\end{equation}
according to the definition of $L^{(k)}_{mnp}$, Eq.~\eq{L-k-mnp-def}. 
Of course, this result is invariant with respect to choice of the coordinate frame.

Finally, the combination of equations~\eq{app-d2chi-dr2-via-omega-eta}, 
\eq{app-int-with-omega}, \eq{app-int-with-eta} provides the following result 
for the integral that appears in Eq.~\eq{app-phi-mnp-alpha-beta-1}:
\begin{multline}
  \iiint f \, \frac{\d^2 \chi}{\d r_\alpha \d r_\beta} \; \frac{d^3r}{|\r-\R|} \\
  = \sum\limits_i n_{i\alpha} \, n_{i\beta} 
  \left(  n_{i\gamma} \Phi^{(i)}_{(mnp),\gamma} + \Omega^{(i)}_{mnp}  \right) 
  + \sum\limits_k \lambda^{(k)}_{\alpha\beta} \, L^{(k)}_{mnp} \, .
\end{multline}
Substituting this expression into Eq.~\eq{app-phi-mnp-alpha-beta-1}, 
one obtains Eq.~\eq{phi-alpha-beta-via-primitives}, which represents the 
quantity $\phi^{(mnp)}_{,\alpha\beta}$ in terms of primitives.

Hence, Eq.~\eq{phi-alpha-beta-via-primitives} is proven, as well as 
Eq.~\eq{phi-alpha-via-phi-Phi}.

\section{Reducing volume integrals to surface integrals}
\label{app:volume-to-surface}

The aim of this Appendix consists in deriving Eq.~\eq{phi-via-Phi}. 
For simplicity, let us consider the case of 
\begin{equation}
  \R = 0 ,
\end{equation}
that does not lead to any loss of generality. (In order to account for nonzero $\R$, 
one can make the substitution $\r \rightarrow \r - \R$.) Let us define a vector field $\vec{F}(\r)$ as
\begin{equation}
  \mathbf{F}(\mathbf{r})  =  \frac{\mathbf{r}}{r} \, x^m y^n z^p ,
\end{equation}
and apply Gauss's theorem:
\begin{equation} \label{Gauss-F}
  \iiint\limits_{\text{inclusion}} (\nabla\cdot\mathbf{F}) \, dV  
  =  \sum\limits_i \iint\limits_{\text{face}\,i} \mathbf{F} \cdot \mathbf{n}_i \, dS .
\end{equation}
The integrand in the left-hand side of Eq.~(\ref{Gauss-F}) is 
\begin{equation}
  \nabla\cdot\mathbf{F} = (m+n+p+2) \frac{x^m y^n z^p}{r} \, .
\end{equation}
Therefore the left-hand side of Eq.~(\ref{Gauss-F}) is equal to $(m+n+p+2)\,\varphi_{mnp}$. 
The integrand in the right-hand side of Eq.~(\ref{Gauss-F}) is
\begin{equation}
  \mathbf{F} \cdot \mathbf{n}_i = \mathbf{r} \cdot \mathbf{n}_i \, \frac{x^m y^n z^p}{r} \, .
\end{equation}
At any point of $i$th face, the factor $\mathbf{r} \cdot \mathbf{n}_i$ has the same value
\begin{equation}
  \mathbf{r} \cdot \mathbf{n}_i = h_i \, ,
\end{equation}
where the quantity $h_i$ is defined in Eq.~\eq{h-i}. Therefore this factor can be 
carried out of the integral, and the integral over $i$th face in Eq.~\eq{Gauss-F} 
becomes equal to $h_i \, \Phi^{(i)}_{mnp}$. Hence, Eq.~(\ref{Gauss-F}) takes the following form:
\begin{equation} \label{app-phi-via-Phi}
  (m+n+p+2)\,\phi_{mnp} = \sum\limits_i h_i \, \Phi^{(i)}_{mnp} \, ,
\end{equation}
that proves validity of Eq.~\eq{phi-via-Phi}.

\section{Converting primitives $\hat\Phi^{(i)}_{mn0}$ to $\hat\Omega^{(i)}_{mn0}$}
\label{app:Phi-to-Omega}

Here we will derive Eq.~\eq{Phi-via-Omega-L}. 
In order to simplify notations, we omit the hats at $\hat x, \hat y, \hat z, \hat r$. 
Consider the two-dimensional vector field $(G_x,G_y)$ whose components are defined as
\begin{equation}
  G_x (x,y) = \frac{x^{m+1} y^n}{r} , \qquad G_y (x,y) = \frac{x^m y^{n+1}}{r} ,
\end{equation}
and write down a two-dimensional version of Gauss's theorem in the plane of $i$th face:
\begin{equation} \label{Gauss-G}
  \iint\limits_{\text{face}\,i} \left( \frac{\d G_x}{\d x} + \frac{\d G_y}{\d y} \right) \, dS  =  
  \sum\limits_{k} 
  \int\limits_{\;\;\text{edge}\,k} (G_x b_{ikx} + G_y b_{iky}) \, dl ,
\end{equation}
where summation is over the edges surrounding face $i$; 
$b_{ikx}$ and $b_{iky}$ are components of the vector $\vec{b}_{ik}$ in the ``tilted'' frame. 

The integrand of the surface integral in Eq.~(\ref{Gauss-G}) can be rewritten as
\begin{multline}
  \frac{\d G_x}{\d x} + \frac{\d G_y}{\d y} 
  = (m+n+1) \, \frac{x^m y^n}{r} + \frac{x^m y^n z^2}{r^3} ,
\end{multline}
and, according to the definitions~\eq{hat-Phi-i-mnp-def} and~\eq{hat-Omega-i-mnp-def}, 
the left-hand side of Eq.~(\ref{Gauss-G}) obtains the following form:
\begin{multline} \label{Gauss-G-left}
  \iint\limits_{\text{face}\,i} \left( \frac{\d G_x}{\d x} + \frac{\d G_y}{\d y} \right) \, dS 
  =  (m+n+1) \, \hat\Phi^{(i)}_{mn0} - \hat\Omega^{(i)}_{mn1} \\
  =  (m+n+1) \, \hat\Phi^{(i)}_{mn0} - h_i \, \hat\Omega^{(i)}_{mn0} \, .
\end{multline}

On the other hand, the integrand of the line integral in Eq.~(\ref{Gauss-G}) is
\begin{equation}
  G_x b_{ikx} + G_y b_{iky}  =  (\mathbf{r}\cdot\mathbf{b}_{ik}) \frac{x^m y^n}{r} \, .
\end{equation}
The factor $\mathbf{r}\cdot\mathbf{b}_{ik}$ is constant along the edge $k$, and has the value
\begin{equation}
  \mathbf{r}\cdot\mathbf{b}_{ik} = B_{ik} 
\end{equation}
due to Eq.~\eq{B-ik}. Thus, 
\begin{equation} \label{Gauss-G-right}
  \int\limits_{\text{edge}\,k} (G_x b_{ikx} + G_y b_{iky}) \, dl  
  =  B_{ik} \int\limits_{\text{edge}\,k} \frac{x^m y^n}{r} \, dl  =  B_{ik} \, \hat L^{(k)}_{mn0} \, .
\end{equation}
Collecting together Eqs.~(\ref{Gauss-G}), (\ref{Gauss-G-left}), and~(\ref{Gauss-G-right}), 
one can get the relation
\begin{equation}
  (m+n+1) \, \hat\Phi^{(i)}_{mn0} - h_i \, \hat\Omega^{(i)}_{mn0}  
  =  \sum\limits_k B_{ik} \, \hat L^{(k)}_{mn0} \, ,
\end{equation}
which is equivalent to Eq.~\eq{Phi-via-Omega-L}.

\section{Decreasing the index $n$ at $\hat\Omega^{(i)}_{mn0}$}
\label{app:decreasing-n}

In this Appendix we will derive the recursive relation~\eq{Omega-reduce-n} 
that allows a step-by-step decrease of the index $n$ at $\hat\Omega^{(i)}_{mn0}$. 
As in the previous Appendix, we omit the hats at $\hat x, \hat y, \hat z, \hat r$.

Let us make simple transformations of the quantity $h_i \hat\Phi^{(i)}_{mn0}$:
\begin{multline} \label{Phi-mn0-1}
  h_i \hat\Phi^{(i)}_{mn0} = \hat\Phi^{(i)}_{mn1} 
  = \iint \frac{x^m y^n z}{r} \, dS \\
  = \iint \frac{x^2+y^2+z^2}{r^2} \, \frac{x^m y^n z}{r} \, dS 
  = \iint \frac{x^{m+2} y^n z}{r^3} \, dS \\ 
  + \iint \frac{x^m y^{n+2} z}{r^3} \, dS + \iint \frac{x^m y^n z^3}{r^3} \, dS \\
  = - \hat\Omega^{(i)}_{m+2,n,0} - \hat\Omega^{(i)}_{m,n+2,0} - \hat\Omega^{(i)}_{mn2} \, .
\end{multline}
(Integrals are over the $i$th face.) On the other hand, one can use 
Eq.~(\ref{Phi-via-Omega-L}) for reducing the value of $h_i \hat\Phi^{(i)}_{mn0}$ 
to quantities $\hat\Omega^{(i)}_{mn0}$ and $\hat L^{(k)}_{mn0}$:
\begin{equation} \label{Phi-mn0-2}
  h_i \hat\Phi^{(i)}_{mn0} = \frac{h_i^2}{m+n+1} \, \hat\Omega^{(i)}_{mn0}  
  +  \sum\limits_k \frac{h_i B_{ik}}{m+n+1} \, \hat L^{(k)}_{mn0} \, ,
\end{equation}
where summation is over edges adjacent to face~$i$. 
Since left-hand sides of Eqs.~(\ref{Phi-mn0-1}) and~(\ref{Phi-mn0-2}) 
are the same, then right-hand sides are equal to each other:
\begin{multline} 
  - \hat\Omega^{(i)}_{m+2,n,0} - \hat\Omega^{(i)}_{m,n+2,0} - \hat\Omega^{(i)}_{mn2}  \\
  = \frac{h_i^2}{m+n+1} \, \hat\Omega^{(i)}_{mn0}  
  +  \sum\limits_k \frac{h_i B_{ik}}{m+n+1} \, \hat L^{(k)}_{mn0} \, .
\end{multline}
Expressing $\hat\Omega^{(i)}_{m,n+2,0}$ from this equation, and taking into account 
that $\hat\Omega^{(i)}_{mn2} = h_i^2 \hat\Omega^{(i)}_{mn0}$ 
due to Eq.~\eq{p-to-zero}, one can get the following result:
\begin{multline} \label{app-Omega-reduce-n}
  \hat\Omega^{(i)}_{m,n+2,0} = - \hat\Omega^{(i)}_{m+2,n,0} \\
  - \frac{m+n+2}{m+n+1} \, h_i^2 \hat\Omega^{(i)}_{mn0} 
  - \sum\limits_k \frac{h_i B_{ik}}{m+n+1} \, \hat L^{(k)}_{mn0} \, .
\end{multline}
Equation~\eq{Omega-reduce-n} can be obtained from 
Eq.~\eq{app-Omega-reduce-n} by means of the substitution $n \to n-2$.

\section{Reducing surface integrals $\hat\Omega^{(i)}_{m10}$ to line integrals}
\label{app:Omega-m10}

In this Appendix, we will derive Eq.~\eq{Omega-m10-answer} that simplifies 
the surface integral $\hat\Omega^{(i)}_{m10}$. 
As above, we will omit here the hats at $\hat x, \hat y, \hat z, \hat r$.

Let us perform integration of $\hat\Omega^{(i)}_{m10}$ over $y$:
\begin{figure}
  \includegraphics[width=0.8\columnwidth]{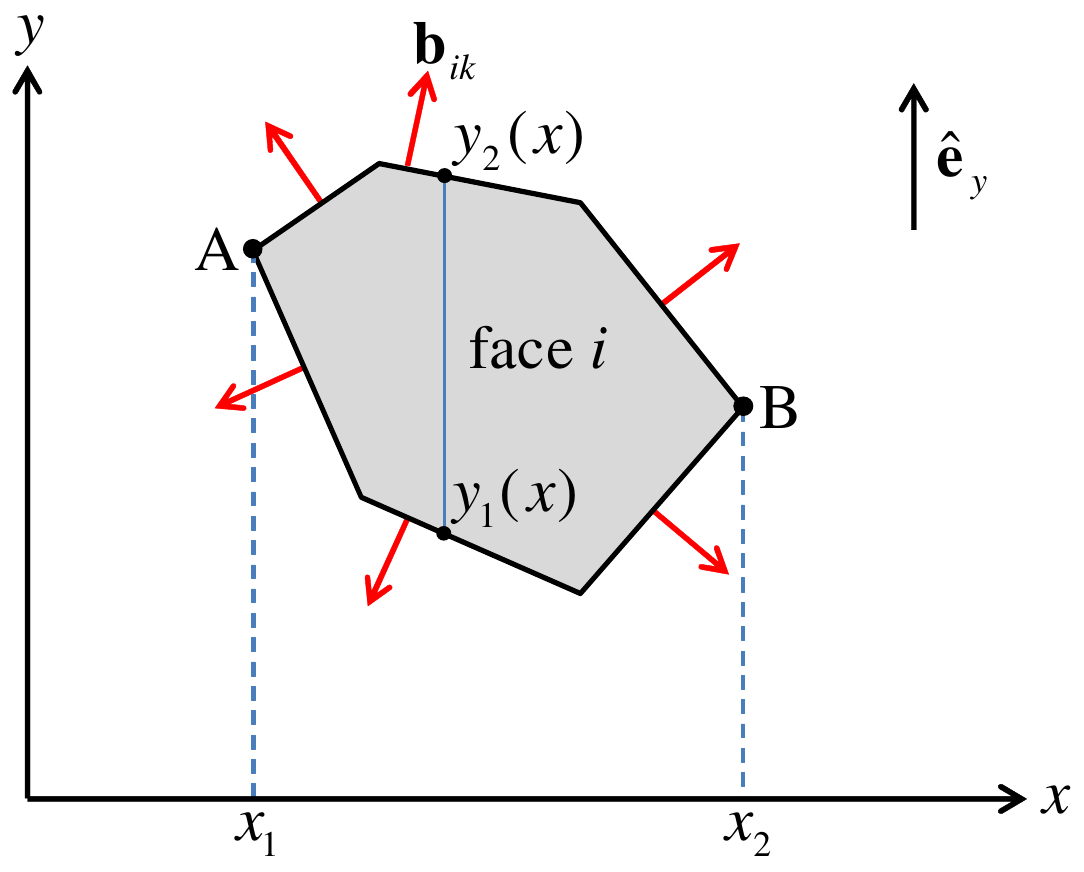}
  \caption{How to evaluate the integral of $\hat\Omega^{(i)}_{m10}$.}
  \label{fig:omega-integration}
\end{figure}
\begin{multline} \label{Omega-m10-calc1}
  \hat\Omega^{(i)}_{m10}
  = - \iint\limits_{\text{face }i} x^m y \frac{z}{r^3} \, dx \, dy 
  = -h_i \int\limits_{x_1}^{x_2} x^m dx \int\limits_{y_1(x)}^{y_2(x)} \frac{y\,dy}{r^3} \\
  =  h_i \int\limits_{x_1}^{x_2} \frac{x^m dx}{r(x,y_2(x))} 
  -  h_i \int\limits_{x_1}^{x_2} \frac{x^m dx}{r(x,y_1(x))} \, ,
\end{multline}
where symbols $x_1,x_2,y_1,y_2$ are explained in Fig.~\ref{fig:omega-integration}; 
$r(x,y) = (x^2+y^2+h_i^2)^{1/2}$; $h_i$ is $z$-coordinate of $i$th face. 
The first integral in the right-hand side of Eq.~(\ref{Omega-m10-calc1}) 
is taken over the upper part $y_2(x)$ of the periphery of $i$th face 
(between points A and B, see Fig.~\ref{fig:omega-integration}), and the second integral 
is over the lower part $y_1(x)$. It is convenient to express $dx$ via 
the line element $dl$ along the edge:
\begin{equation}
  dx = 
  \begin{cases}
    \;\;\; \vec{b}_{ik}\cdot\hat{\vec{e}}_y \, dl   & \text{for upper part}, \\
	     - \vec{b}_{ik}\cdot\hat{\vec{e}}_y \, dl   & \text{for lower part}.
  \end{cases}
\end{equation}
By doing so, one can combine both integrals together, yielding
\begin{equation}
  \hat\Omega^{(i)}_{m10} 
  = h_i \sum\limits_k \vec{b}_{ik}\cdot\hat{\vec{e}}_y 
  \int\limits_{\text{edge}\,k} \frac{x^m}{r} \, dl \, ,
\end{equation}
i. e.
\begin{equation} \label{app-Omega-m10-answer}
  \hat\Omega^{(i)}_{m10} = h_i \sum\limits_k \vec{b}_{ik}\cdot\hat{\vec{e}}_y \, \hat L^{(k)}_{m00} \, .
\end{equation}
Therefore we have obtained Eq.~\eq{Omega-m10-answer}.

\section{Decreasing the index $m$ at $\hat\Omega^{(i)}_{m00}$}
\label{app:decreasing-m}

This Appendix is aimed to deduce the recursive rule~\eq{Omega-reduce-m} 
for reducing the index $m$ at $\hat\Omega^{(i)}_{m00}$. 
As above, we will omit here the hats at $\hat x, \hat y, \hat z, \hat r$.

For this purpose, we rewrite the surface integral $\hat\Omega^{(i)}_{m20}$ in a manner 
similar to Eq.~(\ref{Omega-m10-calc1}):
\begin{multline} \label{Omega-m20-calc1}
  \hat\Omega^{(i)}_{m20}
  = - \iint\limits_{\text{face }i} x^m y^2 \frac{z}{r^3} \, dx \, dy 
  = - h_i \int\limits_{x_1}^{x_2} x^m dx \int\limits_{y_1(x)}^{y_2(x)} \frac{y^2\,dy}{r^3} 
\end{multline}
and make the integration by parts, taking into account that 
$y\,dy = d(r^2)/2$ and $d(y/r^3) = (1/r^3 - 3y^2/r^5)dy$:
\begin{multline}
  \int\limits_{y_1}^{y_2} \frac{y^2\,dy}{r^3} 
  = \int\limits_{y_1}^{y_2} \frac{y}{r^3} \, y\,dy 
  = \left. \frac{y}{r^3} \, \frac{r^2}{2} \right|_{y_1}^{y_2} 
  - \int\limits_{y_1}^{y_2} \frac{r^2}{2} \left( \frac{1}{r^3} - \frac{3y^2}{r^5} \right) dy \\
  = \left. \frac{y}{2r} \right|_{y_1}^{y_2} - \frac12 \int\limits_{y_1}^{y_2} \frac{dy}{r} 
  + \frac32 \int\limits_{y_1}^{y_2} \frac{y^2\,dy}{r^3} \, ,
\end{multline}
whence
\begin{equation}
  \int\limits_{y_1}^{y_2} \frac{y^2\,dy}{r^3}  
  =  \int\limits_{y_1}^{y_2} \frac{dy}{r} - \left. \frac{y}{r} \right|_{y_1}^{y_2}  .
\end{equation}
Substituting this into Eq.~(\ref{Omega-m20-calc1}), one can obtain
\begin{multline} \label{Omega-m20-calc2}
  \hat\Omega^{(i)}_{m20} = 
  - h_i \int\limits_{x_1}^{x_2} x^m dx \int\limits_{y_1(x)}^{y_2(x)} \frac{dy}{r} \\
  + h_i \int\limits_{x_1}^{x_2} \frac{x^m y_2(x)}{r(x,y_2(x))} \, dx
  - h_i \int\limits_{x_1}^{x_2} \frac{x^m y_1(x)}{r(x,y_1(x))} \, dx .
\end{multline}
The first term in the right hand side is equal to $-h_i\hat\Phi^{(i)}_{m00}$ 
(by definition of $\hat\Phi^{(i)}_{m00}$); 
the second and the third terms can be combined together, literally 
repeating the steps that lead from Eq.~(\ref{Omega-m10-calc1}) to Eq.~(\ref{app-Omega-m10-answer}). 
This results in the following representation of $\hat\Omega^{(i)}_{m20}$:
\begin{equation} \label{Omega-m20-answer}
  \hat\Omega^{(i)}_{m20} = 
  - h_i\hat\Phi^{(i)}_{m00} 
  + h_i \sum\limits_k (\vec{b}_{ik}\cdot\hat{\vec{e}}_y) \, L^{(k)}_{m10} \, ,
\end{equation}
where summation is over edges of $i$th face.
The term $(-h_i\hat\Phi^{(i)}_{m00})$ can be reduced to $\Omega$'s by means of Eq.~(\ref{Phi-mn0-1}):
\begin{equation} \label{h-Phi-m00}
  -h_i\hat\Phi^{(i)}_{m00} 
  = \hat\Omega^{(i)}_{m+2,0,0} + \hat\Omega^{(i)}_{m20} + \hat\Omega^{(i)}_{m02} \, .
\end{equation}
Substituting this into Eq.~(\ref{Omega-m20-answer}), 
and expressing $\hat\Omega^{(i)}_{m02}$ as $h_i^2 \hat\Omega^{(i)}_{m00}$ 
according to Eq.~\eq{p-to-zero}, 
one can obtain the following relation:
\begin{equation} \label{app-Omega-reduce-m}
  \hat\Omega^{(i)}_{m+2,0,0} 
  = - h_i^2 \hat\Omega^{(i)}_{m00} 
  - h_i \sum\limits_k (\vec{b}_{ik}\cdot\hat{\vec{e}}_y) \, L^{(k)}_{m10} \, ,
\end{equation}
from which one can get Eq.~\eq{Omega-reduce-m} by means of the substitution $m \rightarrow m-2$.

\section{Recursive formula for line integrals}
\label{app:LL}

Here we will obtain expressions~\eq{L-n-answer}, \eq{L-1-answer} 
for the integrals $\mathcal{L}^{(k)}_1, \mathcal{L}^{(k)}_2 , \ldots$ 
This can be done using integration by parts, 
taking into account that $\xi\,d\xi = \frac12 \, d(\rho_k^2+\xi^2)$:
\begin{multline} \label{L-calc1}
  \mathcal{L}^{(k)}_t
  = \int\limits_{\xi_{k1}}^{\xi_{k2}} \frac{\xi^{t-1}}{\sqrt{\rho_k^2+\xi^2}} \, \xi \, d\xi 
  = \left. \frac{\xi^{t-1}}{\sqrt{\rho_k^2+\xi^2}} \, \frac{\rho_k^2+\xi^2}{2} \right|_{\xi_{k1}}^{\xi_{k2}} \\
  - \frac12 \int\limits_{\xi_{k1}}^{\xi_{k2}}  \frac{ (t-1)\rho_k^2\xi^{t-2} + (t-2)\xi^t }{\sqrt{\rho_k^2+\xi^2}}  \, d\xi .
\end{multline}
According to the definition~\eq{LL}, one can represent the last term 
as a combination of $\mathcal{L}^{(k)}_{t-2}$ and $\mathcal{L}^{(k)}_t$, yielding
\begin{multline} \label{L-calc2}
  \mathcal{L}^{(k)}_t
  = \frac12 \left. \xi^{t-1}\sqrt{\rho_k^2+\xi^2} \right|_{\xi_{k1}}^{\xi_{k2}} 
  - \frac{t-1}{2} \, \rho_k^2 \mathcal{L}^{(k)}_{t-2} - \frac{t-2}{2} \, \mathcal{L}^{(k)}_t \, ,
\end{multline}
whence
\begin{multline} \label{L-calc3}
  t \mathcal{L}^{(k)}_t + (t-1) \rho_k^2 \mathcal{L}^{(k)}_{t-2}
  = \left. \xi^{t-1}\sqrt{\rho_k^2+\xi^2} \right|_{\xi_{k1}}^{\xi_{k2}} \, .
\end{multline}
Square roots $\sqrt{\rho_k^2+\xi^2}$ at $\xi = \xi_{k1}, \xi_{k2}$ have the meaning of 
distances from the point $\R$ to the ends of $k$th edge:
\begin{eqnarray}
  \sqrt{\rho_k^2+\xi_{k1}^2} &=& \left| \r_A^{(k)}-\R \right| ,  \\
  \sqrt{\rho_k^2+\xi_{k2}^2} &=& \left| \r_B^{(k)}-\R \right| .
\end{eqnarray}
Therefore, Eq.~\eq{L-calc3} takes the following form:
\begin{multline} \label{app-L-n-answer}
  t \mathcal{L}^{(k)}_t + (t-1) \rho_k^2 \mathcal{L}^{(k)}_{t-2} \\
  = (\xi_{k2})^{t-1} \left| \r_B^{(k)}-\R \right| 
  - (\xi_{k1})^{t-1} \left| \r_A^{(k)}-\R \right| ,
\end{multline}
Substituting $\xi_{k1}$ and $\xi_{k2}$ from Eq.~\eq{xi-k12} into Eq.~\eq{app-L-n-answer}, 
and resolving the latter equation with respect to $\mathcal{L}^{(k)}_t$, one can get 
the recursive relation~\eq{L-n-answer}.

At $t=1$, Eq.~\eq{app-L-n-answer} is simplified to
\begin{equation} \label{app-L-1-answer}
  \mathcal{L}^{(k)}_1  =  \left| \r_B^{(k)}-\R \right|  -  \left| \r_A^{(k)}-\R \right| .
\end{equation}

\section{Comparison with a closed-form solution found in the literature}
\label{sec:Glas}

In order to check the validity of our analytical results, it is natural to compare them with similar results obtained by other authors. 
But it seems that there are no analytical calculations in the literature, related to polyhedral-shaped inclusions with a smoothly varied 
lattice misfit. For this reason, we choose as a ``benchmark'' the analytical results by Glas\cite{Glas2001} for the strain distribution 
due to an inclusion  having the shape of a truncated pyramid \emph{with a constant misfit strain}. Such an analytical calculations 
can be found in Fig.~3b of Ref.~\onlinecite{Glas2001}. 

\begin{figure*}
  \includegraphics[width=\textwidth]{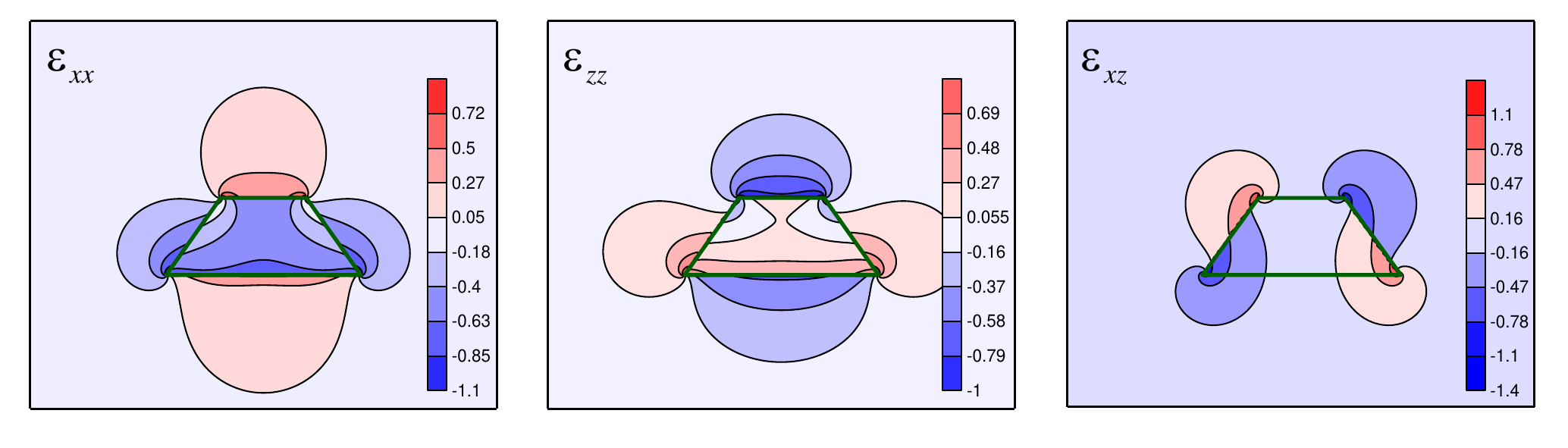}
  \caption{Strain distribution due to an inclusion in the form of a truncated pyramid with a constant misfit strain. 
  This strain distribution is calculated by our method described in Section~VII, 
  and one can see that it exactly reproduces the analytical calculations by Glas, see Fig.~3b in Ref.~\onlinecite{Glas2001}. 
  Top and bottom faces of the truncated pyramid are squares lying in planes (001); four side faces have orientations \{111\}; 
  the length of edges of the bottom face is 5 nm, and the height of the inclusion is equal to 2 nm. Poisson ratio $\nu=1/3$. 
  Strain components $\ve_{xx}$, $\ve_{zz}$ and $\ve_{xz}$ are shown in the plane $y=0$ (a plane of symmetry of the inclusion). 
  The value of the strain is normalized by the misfit strain.}
  \label{fig:Glas}
\end{figure*}

In Fig.~\ref{fig:Glas}, we reproduce these calculations by our method---namely, by means of Eqs.~(82) and~(98). 
For the details of the inclusion geometry, see the caption to the figure. 
One can see that isolines in Fig.~\ref{fig:Glas} are identical to ones in Fig.~3b of Ref.~\onlinecite{Glas2001}. That is, our method 
provides the same strain distribution as the closed-form analytical solution 
found by Glas\cite{Glas2001} for the particular case of a truncated pyramid with a constant misfit strain.

\section{Comparison with numerical calculations}
\label{sec:comsol}

The other useful test is comparison with numerical calculations. One difficulty in the way of such a comparison is that 
numerical methods deal with an elastic body of final size, whereas our analytical approach is valid for an inclusion in an infinitely large matrix. 
Therefore, in order to ensure that the analytical method and is compatible with numerical calculations, one should perform numerical tests for 
different sizes of the matrix and check whether the strain distribution converges to the analytical solution when the matrix size grows.

\begin{figure*}
  \includegraphics[width=\textwidth]{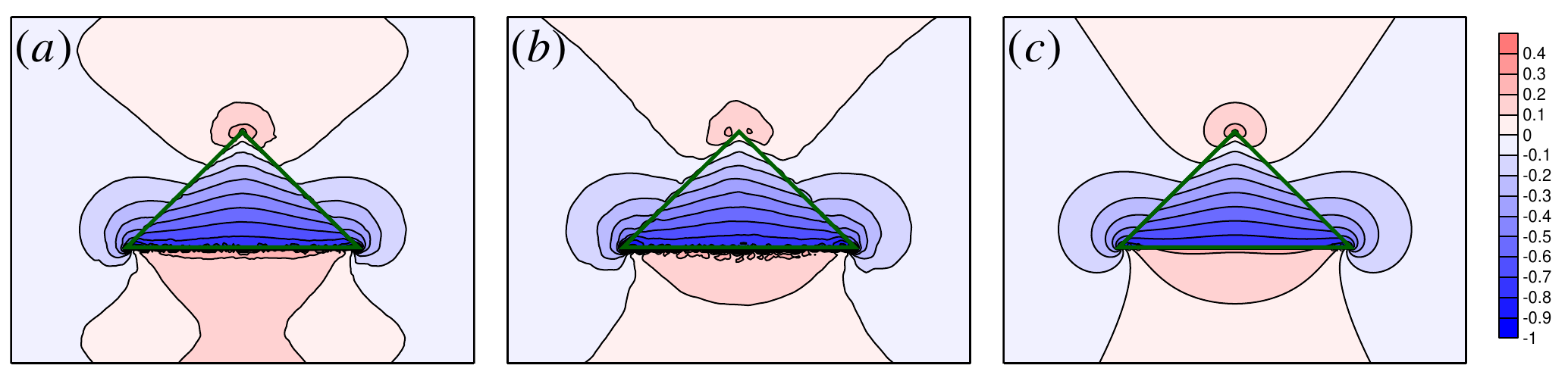}
  \caption{Comparison between numerical and analytical results of the strain calculation. 
  Shown is the distribution of the strain component $\ve_{xx}$ due to a pyramid-shaped inclusion 
  with a misfit strain decreasing linearly from bottom to top: 
  ({\it a}) numerical calculations with the box size $20 \times 20 \times 15$~nm; 
  ({\it b}) numerical calculations with the box size $30 \times 30 \times 25$~nm; 
  ({\it c}) analytical calculations assuming that the inclusion is incorporated in an infinite matrix. 
  One can see that with increasing the box size the numerical results ({\it a,b}) converge to 
  analytical ones ({\it c}). 
  The geometry of the inclusion is the same as in Fig.~8: 
  the pyramid base has orientation (001) and is a square with the side length $l=10$~nm; 
  the pyramid height is $h=5$~nm; the misfit strain near the apex is twice smaller than that near the base. 
  Poisson ratio $\nu=0.25$. 
  The component $\ve_{xx}$ is shown in the plane $y=0$ (a plane of symmetry), the values of the strain are 
  normalized by the misfit strain near the pyramid base. 
  Numerical calculations ({\it a,b}) are performed by the finite-element method using the COMSOL Multiphysics software. 
  Free boundary conditions are applied at the surface of the box. 
  Analytical calculations ({\it c}) are made by the method of Section~VII.}
  \label{fig:comsol}
\end{figure*}

An example of such a comparison is shown in Fig.~\ref{fig:comsol}. As a test sample, a pyramid-shaped inclusion with a linearly varying misfit strain 
is chosen. The parameters of the inclusion (see details in the caption to the figure) are taken the same 
as ones in Fig.~8. Therefore, the analytical solution in Fig.~\ref{fig:comsol}c 
is just the same as shown in Fig.~8. In order to obtain the numerical solutions, we used 
the COMSOL Multiphysics software. Simulations were performed by the finite-difference method for two different sizes of the ``box'', which 
contains the pyramidal inclusion: $20 \times 20 \times 15$~nm (Fig.~\ref{fig:comsol}a) and $30 \times 30 \times 25$~nm (Fig.~\ref{fig:comsol}b), 
whereas the inclusion size is $10 \times 10 \times 5$~nm. One can see that, indeed, with increasing the size of the ``box'', the strain distribution 
converges to the analytical solution. 

It is worth to note that numerical results shown in Fig.~\ref{fig:comsol}a,b demanded about one hour of machine time on PC 
and about 20 gigabytes of memory, whereas the analytical calculation (Fig.~\ref{fig:comsol}c) lasts less than one minute 
and demands less then one megabyte of memory.

\section{Pseudocode representation of the calculation algorithm}
\label{sec:pseudocode}

Calculation of the strain tensor $\ve_{\alpha\beta}(\r)$ for a set of points $\r$ is a two-step procedure. 
The first step is the calculation of polynomials $\mathcal{A}^{(i)}_{\ve\alpha\beta}$, $\mathcal{B}^{(k)}_{\ve\alpha\beta}$, 
$\mathcal{C}^{(s)}_{\ve\alpha\beta}$ for all faces, edges and vertices of the polyhedral inclusion, and for all values of tensor indices 
$\alpha\beta$. The second step is applying Eq.~(82) for each point $\r$, that gives the value of 
$\ve_{\alpha\beta}(\r)$.

Below we represent these steps in a somewhat informal style, as Algorithm~1 and Algorithm~2, respectively. 
Algorithm~1 should be called six times---namely, for each pair $xx, yy, zz, xy, xz$ and $yz$ of the indices $\alpha\beta$. 
Then, Algorithm~2 is called as many times, as many points $\r$ there are, at which the strain tensor $\ve_{\alpha\beta}$ is to be calculated.

Algorithm~2 does not demand any further comments, since it simply calculates $\ve_{\alpha\beta}$ via Eq.~(82). 
But Algorithm~1 surely demands explanations. The idea of Algorithm~1 consists in the fact that, at each stage of the transformations described 
in Sections~II--VI, the function $\ve_{\alpha\beta}(\R)$ is represented as a sum of contributions of 
such functions as $\phi_{mnp}(\R)$, $\Omega^{(i)}_{mnp}(\R)$, etc., each of them being multiplied by some polynomial. Let us denote the polynomial 
corresponding to the function $\phi_{mnp}(\R)$ as $[\phi_{mnp}]$, etc. Therefore,
\[
\ve_{\alpha\beta}(\R) = \sum_{m,n,p}[\phi_{mnp}]\;\phi_{mnp} + \sum_{i,m,n,p}[\Omega^{(i)}_{mnp}]\;\Omega^{(i)}_{mnp} + \ldots
\]
Then, let there be some identity connecting a function $A(\R)$ with other functions $B(\R), \, C(\R)$:
\[
A(\R) = b(\R)\;B(\R) + c(\R)\;C(\R) ,
\]
where $b(\R), \, c(\R)$ are some polynomials. Then, one can rewrite the contribution $[A]\;A$ of the function $A$ into $\ve_{\alpha\beta}$ as
\[
[A]\;A = ([A]\;b)\;B + ([A]\;c)\;C ,
\]
i.~ e. transform it into contributions of $B$ and $C$. Such a transformation can be implemented in the algorithm as a transformation of polynomial $[A]$ 
into polynomials $[B], \; [C]$ as follows:
\begin{algorithmic}[1]
\State $[B] = [A]\;b$
\State $[C] = [A]\;c$
\State $[A] = 0$
\end{algorithmic}
Implementing different identities from Sections~II--VI in a due order, one can transform all the polynomials like 
$[\phi_{mnp}]$, $[\Omega^{(i)}_{mnp}]$, etc. into the following ones: $[\Omega^{(i)}_{000}] \equiv \mathcal{A}^{(i)}_{\ve\alpha\beta}$, 
$[L^{(k)}_{000}] \equiv \mathcal{B}^{(k)}_{\ve\alpha\beta}$, and $[|\R-\r_s|] \equiv \mathcal{C}^{(s)}_{\ve\alpha\beta}$ for different $i$, $k$, $s$. 
This is how one can find out the polynomials $\mathcal{A}$, $\mathcal{B}$, $\mathcal{C}$. This is the idea, on which Algorithm~1 is based.

It is important to note that Algorithm~1 is by no means optimized. Its performance and memory requirements can be easily improved, but we leave it here in its 
non-optimized version, in order to keep its form as close to equations of Sections~II--VI as possible. The numbers of used equations 
are indicated in comments.


\begin{widetext}

{\bf Algorithm 1: finding out the polynomials} $\mathcal{A}^{(i)}_{\ve\alpha\beta}$, $\mathcal{B}^{(k)}_{\ve\alpha\beta}$, 
$\mathcal{C}^{(s)}_{\ve\alpha\beta}$

\ 

\noindent {\bf Input:} 
\newline tensor indices $\alpha, \beta$ determining which strain component $\ve_{\alpha\beta}$ we are going to calculate;
\newline polynomial $C$, which defines the coordinate dependence of the eigenstrain within the inclusion;
\newline degree $N$ of the polynomial $C$;
\newline the geometry of the polyhedral inclusion: coordinates of vertices; the set of unit vectors $\vec{n}_i, \vec{b}_{ik}, \vec{l}_k$ 
that describes orientations of faces and edges; for each face $i$ (edge $k$), one arbitrarily chosen point $\r_i$ ($\r_k$) on it; 
\newline Poisson ratio $\nu$.

\noindent {\bf Output:} polynomials $\mathcal{A}^{(i)}_{\ve\alpha\beta}$ for each face $i$, 
$\mathcal{B}^{(k)}_{\ve\alpha\beta}$ for each edge $k$, 
$\mathcal{C}^{(s)}_{\ve\alpha\beta}$ for each vertex $s$.

\noindent {\bf Variables:}
\newline integer numbers: $i, i_1, i_2, k, m,n,p, m',n',p', t, \gamma$;
\newline real numbers, vectors and arrays: $\Lambda, \lambda^{(k)}_{\alpha\beta}, \hat{\vec{e}}_x, \hat{\vec{e}}_y, \hat{\vec{e}}_z, 
\r_{\mathrm{A}}, \r_{\mathrm{B}}, T^{m'n'p'}_{mnp}$;
\newline polynomials (of degree $N$) and arrays of polynomials: $[\phi_{m,n,p}]$, $[\Phi^{(i)}_{m,n,p}]$, $[\Omega^{(i)}_{m,n,p}]$, 
$[L^{(k)}_{m,n,p}]$, $[\hat\Phi^{(i)}_{m,n,p}]$, $[\hat\Omega^{(i)}_{m,n,p}]$, $[\hat L^{(k)}_{m,n,p}]$, 
$[\phi^{(mnp)}_{\alpha\beta}]$, $[\phi_{(mnp),\alpha\beta}]$, $[\Phi^{(i)}_{(mnp),\gamma}]$, $[\mathcal{L}^{(k)}_t]$, 
$h_i$, $B_{ik}$, $\r_0^{(k)}$, $\rho^2_k$, $\tilde C_{mnp}$, 
$P_{\mathrm{A}}$, $P_{\mathrm{B}}$.
\newline (Note 1: polynomials can be stored as real arrays $(N+1) \times (N+1) \times (N+1)$ of their coefficients.)
\newline (Note 2: this algorithm can be modified for calculating the potential or its first derivatives, 
but in this case some polynomials will have degrees $N+2$ and $N+1$, respectively.)

\noindent\rule[0.5ex]{\linewidth}{1pt}

\begin{algorithmic}[1]

\State \Comment{{\bf Initialization:}}
\State set polynomials $\mathcal{A}^{(i)}_{\ve\alpha\beta}$, $\mathcal{B}^{(k)}_{\ve\alpha\beta}$, 
$\mathcal{C}^{(s)}_{\ve\alpha\beta}$ to zero
\ForAll{$m\ge0, \; n\ge0, \; p\ge0 \;\; (m+n+p\le N)$}
	\State $[\phi_{m,n,p}] = 0$
	\ForAll{faces $i$}
		\State $[\Phi^{(i)}_{m,n,p}] = 0$
	\EndFor
\EndFor
\For{$t = 0$ to $N$}
	\ForAll{edges $k$}
		\State $[\mathcal{L}^{(k)}_t] = 0$
	\EndFor
\EndFor

\ForAll{faces $i$}
	\State define polynomial $h_i$ as follows: $h_i(\R) = (\r_i-\R) \cdot \vec{n}_i$
	\Comment{Eq. (44)}
	\ForAll{edges $k$ adjacent to face $i$}
		\State define polynomial $B_{ik}$ as follows: $B_{ik}(\R) = (\r_k-\R) \cdot \vec{b}_{ik}$
		\Comment{Eq. (45)}
	\EndFor
\EndFor
\ForAll{edges $k$}
	\State define polynomials $\r_0^{(k)} = \big\{ x_0^{(k)}, y_0^{(k)}, z_0^{(k)} \big\}$ and $\rho^2_k$ as follows:
	\State $\r_0^{(k)}(\R) = \r_k - \vec{l}_k \big[ \vec{l}_k \cdot (\r_k-\R) \big]$,
	\Comment{Eq. (73)}
	\State $\rho^2_k(\R) = |\r_k-\R|^2 - \big[ \vec{l}_k \cdot (\r_k-\R) \big]^2$.
	\Comment{Eq. (70)}
\EndFor
\State $\Lambda = \frac{1}{4\pi} \frac{1+\nu}{1-\nu}$ \Comment{Eq. (10)}
\State

\State  \Comment{{\bf Steps 1, 2:}}
\ForAll{$m\ge0, \; n\ge0, \; p\ge0 \;\; (m+n+p\le N)$}
	\State calculate polynomial $\tilde C_{mnp}$ according to Eq.~(19)
	\State $[\phi^{(mnp)}_{\alpha\beta}] = -\Lambda \tilde C_{mnp}$
	\Comment{Eqs. (12), (22)}
\EndFor
\ForAll{faces $i$}
	\State add $\frac{\delta_{\alpha\beta}}{4\pi} C$ to $\mathcal{A}^{(i)}_{\ve\alpha\beta}$
	\Comment{Eqs. (12), (42), (3)}
\EndFor

\State  \Comment{{\bf Step 3:}}
\ForAll{$m\ge0, \; n\ge0, \; p\ge0 \;\; (m+n+p\le N)$}
	\State $[\phi_{(mnp),\alpha\beta}] = [\phi^{(mnp)}_{\alpha\beta}]$
	\Comment{begin of Eq. (34)}
	\ForAll{faces $i$}
		\ForAll{$\gamma\in\{x,y,z\}$}
			\State $[\Phi^{(i)}_{(mnp),\gamma}] = 
			(- n_{i\alpha}\delta_{\beta\gamma} - n_{i\beta}\delta_{\alpha\gamma} + n_{i\alpha}n_{i\beta}n_{i\gamma}) \; [\phi^{(mnp)}_{\alpha\beta}]$
		\EndFor
		\State $[\Omega^{(i)}_{m,n,p}] = n_{i\alpha} n_{i\beta} \; [\phi^{(mnp)}_{\alpha\beta}]$
	\EndFor
	\ForAll{edges $k$}
		\State find two faces $i_1$, $i_2$ adjacent to edge $k$
		\State $\lambda^{(k)}_{\alpha\beta} = n_{i_1\alpha} b_{i_1k\beta} + n_{i_2\alpha} b_{i_2k\beta}$
		\Comment{Eq. (33)}
		\State $[L^{(k)}_{m,n,p}] = \lambda^{(k)}_{\alpha\beta} \; [\phi^{(mnp)}_{\alpha\beta}]$
	\EndFor
	\State $[\phi^{(mnp)}_{\alpha\beta}] = 0$
	\Comment{end of Eq. (34)}
	
	\If    {$(\alpha\beta)=(xx)$ and $m\ge2$} \Comment{begin of Eq. (37)}
		\State add $m(m-1) \; [\phi_{(mnp),\alpha\beta}]$ to $[\phi_{m-2,n,p}]$ 
	\ElsIf {$(\alpha\beta)=(yy)$ and $n\ge2$}
		\State add $n(n-1) \; [\phi_{(mnp),\alpha\beta}]$ to $[\phi_{m,n-2,p}]$ 
	\ElsIf {$(\alpha\beta)=(zz)$ and $p\ge2$}
		\State add $p(p-1) \; [\phi_{(mnp),\alpha\beta}]$ to $[\phi_{m,n,p-2}]$ 
	\ElsIf {$(\alpha\beta)=(xy)$ or $(yx)$ and $m\ge1$ and $n\ge1$}
		\State add $mn \; [\phi_{(mnp),\alpha\beta}]$ to $[\phi_{m-1,n-1,p}]$ 
	\ElsIf {$(\alpha\beta)=(xz)$ or $(zx)$ and $m\ge1$ and $p\ge1$}
		\State add $mp \; [\phi_{(mnp),\alpha\beta}]$ to $[\phi_{m-1,n,p-1}]$ 
	\ElsIf {$(\alpha\beta)=(yz)$ or $(zy)$ and $n\ge1$ and $p\ge1$}
		\State add $np \; [\phi_{(mnp),\alpha\beta}]$ to $[\phi_{m,n-1,p-1}]$ 
	\EndIf
	\State $[\phi_{(mnp),\alpha\beta}] = 0$ \Comment{end of Eq. (37)}
	
	\ForAll{faces $i$}
		\If {$m\ge1$} \Comment{begin of Eq. (38)}
			\State add $m \; [\Phi^{(i)}_{(mnp),x}]$ to $[\Phi^{(i)}_{m-1,n,p}]$ 
		\EndIf
		\If {$n\ge1$}
			\State add $n \; [\Phi^{(i)}_{(mnp),y}]$ to $[\Phi^{(i)}_{m,n-1,p}]$ 
		\EndIf
		\If {$p\ge1$}
			\State add $p \; [\Phi^{(i)}_{(mnp),z}]$ to $[\Phi^{(i)}_{m,n,p-1}]$ 
		\EndIf
		\State $[\Phi^{(i)}_{(mnp),x}] = 0$; $\;\;[\Phi^{(i)}_{(mnp),y}] = 0$; $\;\;[\Phi^{(i)}_{(mnp),z}] = 0$
		\Comment{end of Eq. (38)}
	\EndFor
	
\EndFor

\State  \Comment{{\bf Step 4:}}
\ForAll{$m\ge0, \; n\ge0, \; p\ge0 \;\; (m+n+p\le N)$}
	\ForAll{faces $i$}
		\State add $\frac{1}{m+n+p+2}h_i \; [\phi_{m,n,p}]$ to $[\Phi^{(i)}_{m,n,p}]$  \Comment{begin of Eq. (47)}
	\EndFor
	\State $[\phi_{m,n,p}] = 0$  \Comment{end of Eq. (47)}
\EndFor

\State  
\ForAll{faces $i$}  \Comment{Now we are going to the tilted frame}
\State

\ForAll{$m\ge0, \; n\ge0, \; p\ge0 \;\; (m+n+p\le N)$}
	\State $[\hat\Phi^{(i)}_{m,n,p}] = 0$ \Comment{Initialize polynomials related to the tilted frame}
	\State $[\hat\Omega^{(i)}_{m,n,p}] = 0$
	\ForAll{edges $k$ adjacent to face $i$}
		\State $[\hat L^{(k)}_{m,n,p}] = 0$
	\EndFor
\EndFor

\State  \Comment{{\bf Step 5:}}
\State choose an orthonormal set of vectors $\hat{\vec{e}}_x, \hat{\vec{e}}_y, \hat{\vec{e}}_z$ 
so that $\hat{\vec{e}}_z = \vec{n}_i$
\ForAll{$m\ge0, \; n\ge0, \; p\ge0 \;\; (m+n+p\le N)$}
	\ForAll{$m'\ge0, \; n'\ge0, \; p'\ge0$ such that $m'+n'+p'=m+n+p$}
		\State calculate coefficient $T^{m'n'p'}_{mnp}$ from Eq. (51)
		\State add $T^{m'n'p'}_{mnp} \; [\Phi^{(i)}_{m,n,p}]$ to $[\hat\Phi^{(i)}_{m',n',p'}]$
		\Comment{Eq. (50a)}
		\State add $T^{m'n'p'}_{mnp} \; [\Omega^{(i)}_{m,n,p}]$ to $[\hat\Omega^{(i)}_{m',n',p'}]$
		\Comment{Eq. (50b)}
	\EndFor
	\State $[\Phi^{(i)}_{m,n,p}] = 0$; $\quad [\Omega^{(i)}_{m,n,p}] = 0$ \Comment{Eq. (50)}
\EndFor

\State  \Comment{{\bf Step 6:}}
\ForAll{$m\ge0, \; n\ge0, \; p\ge1 \;\; (m+n+p\le N)$}
	\State add $(h_i)^p \; [\hat\Phi^{(i)}_{m,n,p}]$ to $[\hat\Phi^{(i)}_{m,n,0}]$;
	$\quad [\hat\Phi^{(i)}_{m,n,p}] = 0$
	\Comment{Eq. (53)}
	\State add $(h_i)^p \; [\hat\Omega^{(i)}_{m,n,p}]$ to $[\hat\Omega^{(i)}_{m,n,0}]$;
	$\quad [\hat\Omega^{(i)}_{m,n,p}] = 0$
	\Comment{Eq. (53)}
\EndFor

\State  \Comment{{\bf Step 7:}}
\ForAll{$m\ge0, \; n\ge0 \;\; (m+n\le N)$}
	\State add $\frac{h_i}{m+n+1} \; [\hat\Phi^{(i)}_{m,n,0}]$ to $[\hat\Omega^{(i)}_{m,n,0}]$
	\Comment{begin of Eq. (55)}
	\ForAll{edges $k$ adjacent to face $i$}
		\State add $\frac{B_{ik}}{m+n+1} \; [\hat\Phi^{(i)}_{m,n,0}]$ to $[\hat L^{(k)}_{m,n,0}]$
	\EndFor
	\State $[\hat\Phi^{(i)}_{m,n,0}] = 0$
	\Comment{end of Eq. (55)}
\EndFor

\State  \Comment{{\bf Step 8:}}
\If{$N\ge2$}
	\For{$m=0$ to $N-2$}
		\For{$n=N-m$ down to 2}
			\State add $-[\hat\Omega^{(i)}_{m,n,0}]$ to $[\hat\Omega^{(i)}_{m+2,n-2,0}]$
			\Comment{begin of Eq. (56)}
			\State add $-\frac{m+n}{m+n-1} (h_i)^2 \; [\hat\Omega^{(i)}_{m,n,0}]$ to $[\hat\Omega^{(i)}_{m,n-2,0}]$
			\ForAll{edges $k$ adjacent to face $i$}
				\State add $-\frac{h_i B_{ik}}{m+n-1} \; [\hat\Omega^{(i)}_{m,n,0}]$ to $[\hat L^{(k)}_{m,n-2,0}]$
			\EndFor
			\State $[\hat\Omega^{(i)}_{m,n,0}] = 0$
			\Comment{end of Eq. (56)}
		\EndFor
	\EndFor
\EndIf
\If{$N\ge1$}
	\For{$m=0$ to $N-1$}
		\ForAll{edges $k$ adjacent to face $i$}
			\State add $h_i (\vec{b}_{ik}\cdot\hat{\vec{e}}_y) \; [\hat\Omega^{(i)}_{m,1,0}]$ to $[\hat L^{(k)}_{m,0,0}]$
			\Comment{begin of Eq. (57)}
		\EndFor
		\State $[\hat\Omega^{(i)}_{m,1,0}] = 0$
		\Comment{end of Eq. (57)}
	\EndFor
\EndIf

\State  \Comment{{\bf Step 9:}}
\If{$N\ge2$}
	\For{$n=N$ down to 2}
		\State add $-(h_i)^2 \; [\hat\Omega^{(i)}_{m,0,0}]$ to $[\hat\Omega^{(i)}_{m-2,0,0}]$
		\Comment{begin of Eq. (58)}
		\ForAll{edges $k$ adjacent to face $i$}
			\State add $-h_i (\vec{b}_{ik}\cdot\hat{\vec{e}}_y) \; [\hat\Omega^{(i)}_{m,0,0}]$ to $[\hat L^{(k)}_{m-2,1,0}]$
		\EndFor
		\State $[\hat\Omega^{(i)}_{m,0,0}] = 0$
		\Comment{end of Eq. (58)}
	\EndFor
\EndIf
\If{$N\ge1$}
		\ForAll{edges $k$ adjacent to face $i$}
			\State add $h_i (\vec{b}_{ik}\cdot\hat{\vec{e}}_x) \; [\hat\Omega^{(i)}_{1,0,0}]$ to $[\hat L^{(k)}_{0,0,0}]$
			\Comment{begin of Eq. (59)}
		\EndFor
		\State $[\hat\Omega^{(i)}_{1,0,0}] = 0$
		\Comment{end of Eq. (59)}
\EndIf
\State add $[\hat\Omega^{(i)}_{0,0,0}]$ to $\mathcal{A}^{(i)}_{\ve\alpha\beta}$;
$\quad[\hat\Omega^{(i)}_{0,0,0}] = 0$
\Comment{(since $\hat\Omega^{(i)}_{000} \equiv \Omega^{(i)}$)}

\State
\State  \Comment{{\bf Step 10 in the tilted frame:}}
\ForAll{edges $k$ adjacent to face $i$}
	\ForAll{$m\ge0, \; n\ge0 \;\; (m+n\le N)$}
		\State $p=0$
		\For{$t = 0$ to $m+n+p$}
			\State calculate polynomial $\hat c_{mnp,t}$ according to Eq. (78)
			\State add $\hat c_{mnp,t} \; [\hat L^{(k)}_{m,n,p}]$ to $[\mathcal{L}^{(k)}_t]$
			\Comment{begin of Eq. (77)}
		\EndFor
		\State $[\hat L^{(k)}_{m,n,p}] = 0$
		\Comment{end of Eq. (77)}
	\EndFor
\EndFor

\State
\EndFor \Comment{Return back from the tilted frame}
\State

\State  \Comment{{\bf Step 10 in the non-tilted frame:}}
\ForAll{edges $k$}
	\ForAll{$m\ge0, \; n\ge0, \; p\ge0 \;\; (m+n+p\le N)$}
		\For{$t = 0$ to $m+n+p$}
			\State calculate polynomial $c_{mnp,t}$ according to Eq. (75)
			\State add $c_{mnp,t} \; [L^{(k)}_{m,n,p}]$ to $[\mathcal{L}^{(k)}_t]$
			\Comment{begin of Eq. (76)}
		\EndFor
		\State $[L^{(k)}_{m,n,p}] = 0$
		\Comment{end of Eq. (76)}
	\EndFor
\EndFor

\State  \Comment{{\bf Step 11:}}
\ForAll{edges $k$}
	\State Let A and B be numbers of the two vertices connected by edge $k$, and unit vector 
	$\vec{l}_k$ be directed from point A to point B. Let $\r_{\mathrm{A}}$ and $\r_{\mathrm{B}}$ 
	be position vectors of these points.
	\State Define two polynomials $P_{\mathrm{A}}$ and $P_{\mathrm{B}}$ as follows:
	\State $P_{\mathrm{A}}(\R) = \vec{l}_k \cdot (\r_{\mathrm{A}}-\R)$, \quad
	$P_{\mathrm{B}}(\R) = \vec{l}_k \cdot (\r_{\mathrm{B}}-\R)$.
	\If{$N \ge 2$}
		\For{$t=N$ down to 2}
			\State add $-\frac{t-1}{t} \rho^2_k \; [\mathcal{L}^{(k)}_t]$ to $[\mathcal{L}^{(k)}_{t-2}]$
			\Comment{begin of Eq. (80)}
			\State add $\frac{(P_{\mathrm{B}})^{t-1}}{t} \; [\mathcal{L}^{(k)}_t]$ to $\mathcal{C}^{(\mathrm{B})}_{\ve\alpha\beta}$
			\State add $-\frac{(P_{\mathrm{A}})^{t-1}}{t} \; [\mathcal{L}^{(k)}_t]$ to $\mathcal{C}^{(\mathrm{A})}_{\ve\alpha\beta}$
			\State $[\mathcal{L}^{(k)}_t] = 0$
			\Comment{end of Eq. (80)}
		\EndFor
	\EndIf
	\If{$N \ge 1$}
		\State add $[\mathcal{L}^{(k)}_1]$ to $\mathcal{C}^{(\mathrm{B})}_{\ve\alpha\beta}$
		\Comment{begin of Eq. (81)}
		\State add $-[\mathcal{L}^{(k)}_1]$ to $\mathcal{C}^{(\mathrm{A})}_{\ve\alpha\beta}$
		\State $[\mathcal{L}^{(k)}_1] = 0$
		\Comment{end of Eq. (81)}
	\EndIf
	\State add $[\mathcal{L}^{(k)}_0]$ to $\mathcal{B}^{(k)}_{\ve\alpha\beta}$;
	$\quad[\mathcal{L}^{(k)}_0] = 0$
	\Comment{(since $\mathcal{L}^{(k)}_0 \equiv L^{(k)}$)}
\EndFor

\end{algorithmic}

\noindent\rule[0.5ex]{\linewidth}{1pt}


\ 

\

{\bf Algorithm 2: calculating the strain tensor $\ve_{\alpha\beta}(\r)$ at one point $\r$}

\ 

\noindent {\bf Input:} 
\newline coordinates of the point $\r$ where the strain tensor is to be calculated;
\newline coordinates of vertices of the polyhedral inclusion, connections between vertices, edges and faces of the polyhedron;
\newline the set of polynomials $\mathcal{A}^{(i)}_{\ve\alpha\beta}$, $\mathcal{B}^{(k)}_{\ve\alpha\beta}$, 
$\mathcal{C}^{(s)}_{\ve\alpha\beta}$ for all faces, edges and vertices of the polyhedron (as calculated by Algorithm~1).

\noindent {\bf Output:} the value of the strain tensor $\ve_{\alpha\beta}$ at point $\r$.

\noindent {\bf Variables:}
\newline integer numbers: $i, k, s, \alpha, \beta$;
\newline real numbers: $A, B, C, \Omega, L, R, l_A, l_B, l_{AB}$.

\noindent\rule[0.5ex]{\linewidth}{1pt}

\begin{algorithmic}[1]

\For{$\alpha\in\{x,y,z\}$}
	\For{$\beta\in\{x,y,z\}$}
		\State $\ve_{\alpha\beta} = 0$
	\EndFor
\EndFor

\State
\ForAll{faces $i$} \Comment{the first sum in rhs of Eq. (82)}
	\State Calculate the signed solid angle $\Omega$ subtended by face $i$ at point $\r$. 
	\State (The sign of $\Omega$ is positive (negative) when the outer (the inner) side of face~$i$ is seen from point~$\r$.)
	\State \Comment{Recipes for calculating solid angles can be found in Refs.~\onlinecite{Nenashev2007arxiv, Werner1996}.}
	\ForAll{pairs $\alpha\beta \in \{xx,yy,zz,xy,xz,yz\}$}
		\State $A = \mathcal{A}^{(i)}_{\ve\alpha\beta}(\r)$
		\Comment{here we calculate the value of the polynomial function $\mathcal{A}^{(i)}_{\ve\alpha\beta}(x,y,z)$ at point $\r$}
		\State $\ve_{\alpha\beta} = \ve_{\alpha\beta} + A \Omega$
	\EndFor
\EndFor

\State
\ForAll{edges $k$} \Comment{the second sum in rhs of Eq. (82)}
	\State Let $\r_A$ and $\r_B$ be radius vectors of the two ends of edge $k$.
	\State $l_A = |\r-\r_A|$;   $\;\;l_B = |\r-\r_B|$;   $\;\;l_{AB} = |\r_A-\r_B|$
	\State $L = \log \frac{l_A+l_B+l_{AB}}{l_A+l_B-l_{AB}}$ \Comment{Eq. (43)}
	\ForAll{pairs $\alpha\beta \in \{xx,yy,zz,xy,xz,yz\}$}
		\State $B = \mathcal{B}^{(k)}_{\ve\alpha\beta}(\r)$
		\State $\ve_{\alpha\beta} = \ve_{\alpha\beta} + B L$
	\EndFor
\EndFor

\State
\ForAll{vertices $s$} \Comment{the third sum in rhs of Eq. (82)}
	\State Let $\r_s$ be the radius vector of vertex $s$.
	\State $R = |\r-\r_s|$;
	\ForAll{pairs $\alpha\beta \in \{xx,yy,zz,xy,xz,yz\}$}
		\State $C = \mathcal{C}^{(s)}_{\ve\alpha\beta}(\r)$
		\State $\ve_{\alpha\beta} = \ve_{\alpha\beta} + C R$
	\EndFor
\EndFor

\State
\State $\ve_{yx} = \ve_{xy}$;    $\quad \ve_{zx} = \ve_{xz}$;    $\quad \ve_{zy} = \ve_{yz}$

\end{algorithmic}

\noindent\rule[0.5ex]{\linewidth}{1pt}

\end{widetext}

\bibliography{elastic}
\end{document}